\documentclass[reprint,superscriptaddress,amsmath,amssymb,aps,prx]{revtex4-2}

\usepackage{placeins}
\usepackage{graphicx}
\usepackage{dcolumn}
\usepackage{bm}
\usepackage{xcolor}
\usepackage{booktabs}
\usepackage{tabularx}
\usepackage{longtable}
\usepackage{hyperref}
\usepackage [capitalize] {cleveref}

\begin{document}

\title{Data-Driven Approach to Model the Influence of Magnetic Geometry in the Confinement of Fusion Devices}

\author{R. Laia}
\email{rodrigo.laia@tecnico.ulisboa.pt}
\affiliation{Instituto de Plasmas e Fusão Nuclear, Instituto Superior Técnico, Universidade de Lisboa, 1049-001 Lisboa, Portugal}
\author{R. Jorge}
\affiliation{Instituto de Plasmas e Fusão Nuclear, Instituto Superior Técnico, Universidade de Lisboa, 1049-001 Lisboa, Portugal}
\affiliation{Department of Physics, University of Wisconsin-Madison, Madison, Wisconsin 53706, USA}
\author{G. Abreu}
\affiliation{Alfredo AI, 1600-171 Lisboa, Portugal}

\begin{abstract}
The design of fusion energy devices involves a balance between competing performance metrics to achieve an energy gain. In stellarators, the geometry is very flexible and involves a large number of free parameters. These can be optimized to achieve good performance. One of the main optimization targets is omnigenity, that is, the confinement of alpha particles stemming from the fusion reactions. In this work, two classes of omnigenous stellarators are studied, namely quasisymmetric and quasi-isodynamic stellarators. The goal is to determine the influence of the geometry on omnigenity, which can lead to greater insight into the design space of stellarators. For this purpose, a database of stellarator configurations is created and analyzed for correlations, pair-wise distributions, and dimensionality reduction using a supervised autoencoder framework. Then, a classification model is trained on this database to predict the convergence of numerical solvers. Finally, two regression models, LightGBM and its probabilistic version, LightGBM LSS, as well as a feed-forward neural network, are trained to predict quasisymmetry and quasi-isodynamiticity and find the design parameters that most influence omnigenity.
\end{abstract}


\maketitle

\section{Introduction}
\begin{figure*}
\centering
\includegraphics[trim=2.6cm 4.0cm 2.3cm 4.0cm, clip, width=0.3\linewidth]{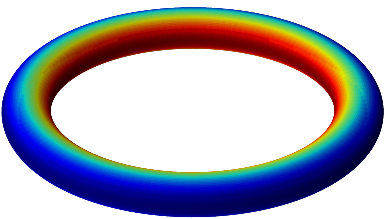}
\includegraphics[trim=2.6cm 3.9cm 2.3cm 4.2cm, clip, width=0.3\linewidth]{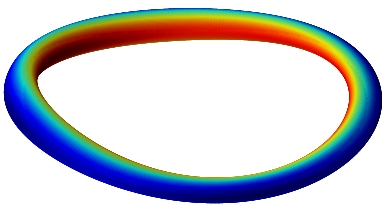}
\includegraphics[trim=2.4cm 3.7cm 2.2cm 3.8cm, clip, width=0.3\linewidth]{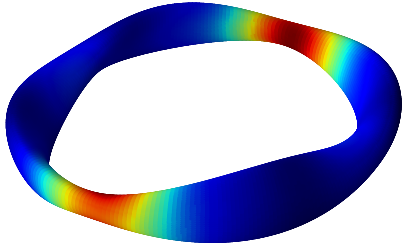}
\caption{Comparison between a tokamak (left), a quasisymmetric (center), and a quasi-isodynamic stellarator (right). Colors show the magnetic field strength at the plasma boundary. All configurations have an aspect ratio of 7, and the stellarators have 2 field periods.}
\label{fig:tokamak_vs_stel}
\end{figure*}

One of the main challenges in fusion energy is to maintain heat and particle transport at sustainable levels while achieving optimal energy confinement.
Such a transport is dependent on the particular geometry used for the device.
Two of the leading candidates in magnetic confinement fusion energy are tokamaks, \cref{fig:tokamak_vs_stel} (left), and stellarators, \cref{fig:tokamak_vs_stel} (center and right). 
Tokamaks have an axisymmetric shape with a central solenoid that drives an internal plasma current to twist the magnetic field and minimize large radial particle drifts.
However, this current is also a source of free energy that can lead to disruptions \cite{Schuller1995DisruptionsTokamaks}.
Stellarators, on the other hand, twist the magnetic field without relying on a central solenoid or toroidally driven current \cite{Xu2016}.
In contrast with tokamaks, stellarators allow steady-state operation but require complex coil configurations to achieve the desired magnetic fields \cite{Helander2014}.

The optimization of stellarator magnetic fields, whether using a fixed-boundary approach \cite{Landreman2021b,Landreman2022}, a single-stage approach \cite{Jorge2023,Jorge2024}, or other combined plasma-coil algorithms \cite{Henneberg2021}, is typically performed by varying a toroidal surface of constant magnetic flux $\psi$ (such as the ones in \cref{fig:tokamak_vs_stel}).
This surface is used as a boundary condition to solve the ideal magnetohydrodynamic (MHD) equation $\mathbf J \times \mathbf B = \nabla p$ where $\mathbf J = \nabla \times \mathbf B/\mu_0$ is the plasma current, $\mathbf B$ the equilibrium magnetic field, and $p$ the plasma pressure.
By obtaining $\mathbf B$, important performance parameters can be obtained, such as plasma volume, MHD stability, neoclassical and turbulent transport.
Here, we solve the ideal MHD equation using the VMEC code \cite{Hirshman1983}.

Both tokamak and stellarator magnetic fields need to be carefully designed to achieve good performance.
One of the challenges of non-axisymmetric geometries, such as the stellarator, is the confinement of the alpha particles stemming from the fusion reactions.
Stellarators with such a property are called omnigenous, with the two main types of omnigenity being quasisymmetry \cite{Boozer1983,Nuhrenberg1988,Landreman2022} and quasi-isodynamicity \cite{Jorge2022a,Goodman2023}.
The proxies used today to evaluate omnigenity are based on the evaluation of the time averaged radial particle drift $\Delta \psi$, which, in omnigenity, should vanish, that is $\Delta \psi = \int \mathbf v_D \cdot \nabla \psi dt=0$, where $\psi$ is used as a radial coordinate, $\mathbf v_D$ is the particle drift comprised of the grad-B and curvature drifts, $\nabla \psi$ is the vector pointing in the radial direction (outward), and $t$ is time.
Owing to the fact that particle drifts can be simplified using field-aligned coordinates \cite{Boozer1983}, $\Delta\psi=0$ can be translated into a condition dependent on the equilibrium magnetic field only \cite{Plunk2019}.
In quasisymmetry, such condition can be written as $B=B(\psi, M \theta-N\phi)$ with $(\psi,\theta,\phi)$ field-aligned coordinates entitled Boozer coordinates \cite{Boozer1983}, $M$ and $N$ integers, $\theta$ a poloidal angle, $\phi$ a toroidal angle, and $B=|\mathbf B|$.
In quasi-isodynamicity, $B$ contours at constant $\psi$ close poloidally, but $B$ is still dependent on the three variables $B(\psi,\theta,\phi)$, with the added constraint that points of equal magnetic field strength must be parallel \cite{Plunk2019}.

Recent stellarator optimization efforts have shown that precise omnigenity can be achieved through numerical optimization \cite{Landreman2022, Goodman2023}.
In this case, loss from particle drifts can be minimized to values lower than those in tokamaks.
However, optimization for precise omnigenity can be highly dependent on the number of parameters used to describe the stellarator shape and on the initial condition used for the optimization.

In this work, we intend to study the influence of the shape of the stellarator on the relative level of quasisymmetry and quasi-isodynamicity.
Such an insight is crucial to creating cost-effective ways to design stellarators.
Today's precise omnigenous designs are based on gradient-based optimizers such as SIMSOPT \cite{Landreman2021b} and DESC \cite{Dudt2020}, which work by either finite-differencing an ideal MHD solver or using auto-differentiation.
We intend to quantify the role of each surface degree of freedom in the loss functions used to find precisely omnigenous designs, as well as create a quick tool to generate such designs.

For this purpose, we create a database of stellarator configurations by random sampling possible boundary shapes.
We narrow down the search space to stellarators in vacuum with two field periods, major radius of $1$ m, and elliptical cross-sections.
This yields a total of eight degrees of freedom.
As we show here, these are enough degrees of freedom to obtain omnigenous configurations according to standard metrics used in the literature.
However, we note that, in order to achieve precise quasisymmetry or precise quasi-isodynamic conditions, more Fourier modes on the boundary would need to be used.
While stellarator databases have been reported in the literature, these were either obtained in the high aspect ratio and quasisymmetry approximation \cite{Curvo2025UsingDevices,Landreman2022MappingExpansion} or focusing only on a single physics metric \cite{Landreman2025}.
The database created here employs a full MHD model, models both quasisymmetry and quasi-isodinamicity, as well as other important physics metrics.
Furthermore, the analysis presented here allows for the understanding of correlations, clustering, and feature importance on these metrics.

We perform several studies in the database in order to gain further insight into the space of omnigenous stellarators.
This includes algorithms for outlier detection, marginal distributions using Kernel Density Estimation, correlation maps, Principal Component Analysis, and supervised autoencoders for dimensionality reduction.
Additionally, we identify the most influential features affecting these stellarator properties using a LightGBM (Light Gradient Boosting Machine) \cite{Ke2017} classifier for binary classification, along with both LightGBM and LightGBM LSS (Location, Scale, and Shape) \cite{Marz2022} regression models for continuous predictions.

Beyond gradient boosting models, we also show that feed-forward neural networks (FFNN) \cite{Murat2006} can predict quasisymmetry and quasi-isodynamic properties.
The neural network models are trained and evaluated to assess their predictive capabilities and to identify the most influential features affecting these stellarator properties. 

One of the methods used to evaluate the influence of the boundary features on stellarator properties is using the concept of SHAP values \cite{Shapley1971CoresGames,Lipovetsky2001AnalysisApproach}.
This is a concept that originated from game theory, and is a fair way to divide value among the input features, based on how the model's performance degrades when that feature is removed, and considering all subsets of features.
The formulation used here is the one based on the Shapley Additive exPlanation (SHAP) values used in the SHAP Python package \cite{Lundberg2017APredictions}.
Such a formulation has been used recently in Ref. \cite{Landreman2025} to analyze the impact of geometric quantities on the turbulent transport of stellarator devices caused by the ion temperature gradient.
We point out that a recent database using QI-only ideal MHD equilibria has been published in Ref. \cite{Cadena2025}, where a Gaussian Mixture Model \cite{Curvo2025UsingDevices}, together with a Principal Component Analysis (PCA) and Random Forest classifiers, were used to create a generative model that provides stellarator equilibria without running an ideal MHD equilibrium solver such as VMEC.

This paper is organized as follows.
In \cref{sec:selection_database}, we introduce the fundamentals of magnetic field equilibrium, with an emphasis on the derivation of the ideal MHD equations and the Fourier coefficients that form the foundation of the database and models used to study quasisymmetric and quasi-isodynamic configurations. We also describe the stellarator database structure, focusing on how it supports model training and optimization
The analysis of the database entries, including correlations, dimensionality reduction techniques and visualization of distribution densities, is performed in \cref{sec:database_properties}.
In \cref{section:MLmodels}, the gradient boosting tree models to predict quasisymmetry and quasi-isodynamic are described, and in \cref{sec:gbm}, the results from the models are presented.
\cref{section:FFNN} focuses on feed-forward neural networks and their comparison with gradient boosting models.
The conclusions follow.
The code developed for this study is openly accessible on GitHub \cite{Laia2025}, and the database is available on Zenodo \cite{Laia2024}.

\section{Selection of Database Parameters}
\label{sec:selection_database}

In this section, we describe the physical model used to construct the magnetic field equilibria, as well as the performance parameters extracted from such equilibria.
We also describe the database created with these parameters and their range.
Such a database will serve as the basis for the models developed in the next sections.

At the temperatures needed to generate energy from fusion reactions, the matter inside the fusion device becomes a plasma.
Plasmas can be modeled in a myriad of ways, depending on the specific conditions they face and the corresponding approximations that can be used there.
Here, we will use the static version of the ideal MHD equation, $\mathbf J \times \mathbf B = \nabla p$, to find the equilibrium magnetic field $\mathbf B$ able to sustain a plasma pressure $p$ and plasma current density $J$ \cite{Freidberg2014}.
This formulation assumes that typical timescales are larger than collisional time scales and Larmor gyration time scales, and that typical spatial scales are longer than Larmor orbits.
This is the case here as we are interested in the equilibrium magnetic field that confines the plasma at energy confinement time scales, as opposed to plasma fluctuations at a finer level, which can be modelled in the collisional case using two-fluid models \cite{Braginskii1965} or gyrokinetic models \cite{Brizard2007a}.

In ideal MHD, surfaces of constant pressure $p=p(\psi)$ with $\psi$ the toroidal magnetic field flux, are called flux surfaces.
The last closed flux surface is called the plasma boundary.
Such a boundary will be used here as a boundary condition for the ideal MHD model.
Both the magnetic field and plasma current align with flux surfaces as they satisfy in the ideal MHD model $\mathbf B \cdot \mathbf n = \mathbf J \cdot \mathbf n = 0$ with $\mathbf n = \nabla \psi/|\nabla \psi|$ the unit normal vector from the surface and $\nabla p = p'(\psi)\nabla \psi$.
As dictated by the Poincaré index theorem, such surfaces should possess toroidal geometry \cite{Helander2014}, similarly to what is shown in \cref{fig:tokamak_vs_stel}, which will be the geometry used here.

While the ideal MHD model does not guarantee a set of nested toroidal magnetic flux surfaces in general geometries, in this work, we use the VMEC code \cite{Hirshman1983}, which solves the ideal MHD equation assuming a toroidal domain for the equilibrium solution and nested flux surfaces.
VMEC finds this equilibrium by minimizing the potential energy $W$ using the steepest descent method
\begin{equation}
W = \int_V \left(\frac{|\mathbf{B}|^2}{2\mu_0} + \frac{p}{\gamma - 1}\right) d^3x,
\end{equation}
\noindent where \(\gamma\) is the adiabatic index. The outermost flux surface serves as a boundary condition and is represented by a Fourier series in cylindrical coordinates $(R, \phi, z)$, as
\begin{equation}
    R(\theta, \phi) = \sum_{m=0}^{\text{mpol}} \sum_{n=-\text{ntor}}^{\text{ntor}} \text{RBC}_{m,n} \cos(m\theta - n_{\text{fp}}n\phi),
\end{equation}
\begin{equation}
    Z(\theta, \phi) = \sum_{m=0}^{\text{mpol}} \sum_{n=-\text{ntor}}^{\text{ntor}} \text{ZBS}_{m,n} \sin(m\theta - n_{\text{fp}}n\phi).
\end{equation}
\noindent where \(\text{RBC}_{m,n}\) and \(\text{ZBS}_{m,n}\) are the Fourier coefficients of the plasma boundary, \(n_{\text{fp}}\) is the number of field periods, \(\theta \in [0,2\pi[\) is a poloidal angle, and \(\phi \in [0,2\pi[\) is the standard cylindrical angle.

\begin{figure*}
    \centering
    \includegraphics[width=.8\textwidth]{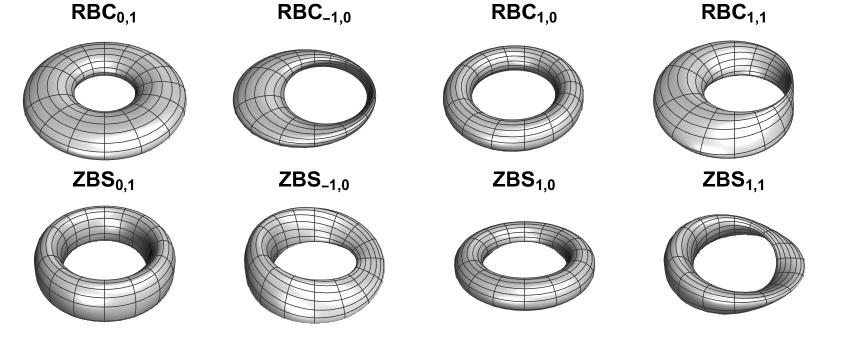}
    \caption{A set of eight stellarators where a constant value of 0.2 is added to the array $\mathbf x$ to a circular torus with $\text{RBC}_{1,0}=\text{ZBS}_{1,0}=0.25$ and the remaining ones equal to zero. This allows us to visually perceive the effect of each degree of freedom in the resulting stellarator shape.}
    \label{fig:stellarator_grid}
\end{figure*}

In this work, we fix $\text{mpol}=\text{ntor}=1$ leading to elliptical shapes in the plane perpendicular to the magnetic axis, we fix $n_{\text{fp}}=2$ to focus on configurations with two field periods, and set the major radius to be one meter using $\text{RBC}_{0,0}=1$.
Therefore, our degrees of freedom are
\begin{align}
    \mathbf x = &[\text{RBC}_{1,0},\text{RBC}_{-1,1},\text{RBC}_{0,1},\text{RBC}_{1,1},\nonumber\\&
    \text{ZBS}_{1,0},\text{ZBS}_{-1,1},\text{ZBS}_{0,1},\text{ZBS}_{1,1}],
\label{eq:dofs}
\end{align}
which is a total of eight independent parameters to specify the plasma boundary and, therefore, the confining magnetic field by solving ideal MHD.
In order to illustrate the effect of each degree of freedom of the surface on the shape of the surface, we show in \cref{fig:stellarator_grid} a set of eight stellarators where a constant value of 0.2 is added to the array $\mathbf x$ to a circular torus with $\text{RBC}_{1,0}=\text{ZBS}_{1,0}=0.25$ and the remaining ones equal to zero.

While a magnetic field equilibrium can be found with a set of nested flux surfaces, due to its non-axisymmetric character, it may lead to large neoclassical transport and loss of alpha particles that stem from fusion reactions.
Such a transport is directly related to the long radial drift of particles $\Delta \psi$.
Omnigenous stellarators, that is, the ones with $\Delta\psi=0$, are usually found by computing proxies of $\Delta\psi$ based on geometry, as opposed to tracing particles for a long time and counting how many are lost to the wall.
For this purpose, we compute commonly used metrics for two types of omnigenity: quasisymmetric and quasi-isodynamic magnetic fields.

Quasisymmetry is a symmetry in the toroidal magnetic field strength $B=|\mathbf B|$ which, when written using Boozer coordinates $(\psi,\theta,\phi)$, depend on the Boozer angles $(\theta,\phi)$ on a surface $\psi=$constant via the combination $M\theta-N\phi$ only, that is, $B=B(\psi,M\theta-N\phi)$ with $M$ and $N$ integers.
This yields a cyclic coordinate in the guiding center Lagrangian, leading to the conservation of canonical angular momentum and, therefore, confining charge particles \cite{Jorge2020a}.
Given that we obtain $\mathbf B$ using a cylindrical coordinate system in VMEC as opposed to Boozer coordinates, we use the following proxy for quasisymmetry $f_{QS}$ \cite{Landreman2022}
\begin{align}
    f_{\text{QS}} = &\sum_{s_j}\left<\left(\frac{1}{B^3}\left[(N-\iota M)\mathbf B \times \nabla B \cdot \nabla \psi\right.\right.\right.\nonumber\\
    &\left.\left.\left.-(MG+NI)\mathbf B \cdot \nabla B\right]\right)^2\right>,
\label{eq:fqs}
\end{align}
where $G(\psi)$ is $\mu_0/(2\pi)$ times the poloidal current outside the surface, $I(\psi)$ is $\mu_0/(2\pi)$ times the toroidal current inside the surface, $\iota$ is the rotational transform and $\left<\dots\right>$ is a flux surface average.
The sum is over a set of flux surfaces $s_j=\psi_j/\psi_b$ where $\psi_b$ is the toroidal flux at the boundary, and a uniform grid $0, 0.1, \dots, 1$ is used.
The quantities $\mathbf B \times \nabla B \cdot \nabla \psi$, $\mathbf B \cdot \nabla B$, $B$, $G$ and $I$ are computed using VMEC.
Given the fact that we focus on $n_{\text{fp}}=2$ configurations, where quasisymmetry is expected to yield mostly quasi-axisymmetric (QA) configurations, we set $M=1$ and $N=0$.
Other options are quasi-poloidal (QP) configurations with $M=0$ and $N=1$ and quasi-helically (QH) symmetric configurations with $M=1$ and $N=1$.
However, quasi-poloidal symmetric is not, in general, allowed close to the magnetic axis \cite{Plunk2019}, and quasi-helical symmetry requires large axis excursions, which typically occur at $n_{\text{fp}}=3$ or larger \cite{Landreman2022MappingExpansion}.
Given the fact that quasisymmetry is a symmetry on the magnetic field strength $B$, the three flavors can also be distinguished by the direction of closed contours of constant magnetic field $B$ on the surface of constant $\psi$.
In this case, QA is said to close toroidally, QP poloidally, and QH helically.

Quasi-isodynamic fields, on the other hand, are omnigenous magnetic fields where the contours of constant $B$ close poloidally but are not necessarily straight, that is, $B$ can have a dependence on both $\phi$ and $\theta$.
We follow the approach in \cite{Goodman2022} and use the following proxy for quasi-isodynamic fields
\begin{equation}
    f_\text{QI}=\frac{n_\text{fp}}{4 \pi^2}\sum_{s_j}\int_0^{2\pi}d\alpha\int_0^{2\pi/n_{\text{fp}}}d\varphi\left(\frac{B-B_{QI}}{B_{\text{max}}-B_{\text{min}}}\right)^2,
\end{equation}
where $B_{QI}$ is the target magnetic field, $\alpha$ is a fieldline label that in Boozer coordinates can be written as $\alpha=\vartheta-\iota \varphi$, and $B_\text{max}$ ($B_\text{min}$) is the maximum (minimum) magnetic field strength on a flux surface.
As QI-optimized magnetic fields often result in elongated flux surfaces and large differences between the maximum and minimum magnetic field strengths on a given flux surface, we also consider the effective elongation $\ell$ as defined in \cite{Goodman2022,Jorge2023} and the mirror ratio $\Delta$ defined as
\begin{equation}
    \Delta = \frac{B_\text{max}-B_\text{min}}{B_\text{max}+B_\text{min}},
\end{equation}
in our characterization of stellarator shapes.
While quasi-axisymmetry is considered to have less shaping than quasi-isodynamic devices, potentially leading to simpler coils, quasi-isodynamic fields are appealing as they can be shown to minimize bootstrap current \cite{Goodman2023}, being beneficial for avoiding current-driven instabilities and facilitating island-divertor operation.

Besides omnigenity, we consider in this work the following physical quantities that are also used to assess the overall properties of a stellarator device: mirror ratio $\Delta$, elongation $\ell$, rotational transform $\iota$, inverse aspect ratio $\epsilon$, magnetic shear $\iota'(\psi)$ and magnetic well $V''(\psi)$.
Larger mirror ratios increase the maximum magnetic field, leading to more strain in the coils, and decrease the minimum magnetic field, yielding poorer confinement, so that smaller values of $\Delta$ are preferred.
Similarly, smaller values of elongation are preferred as larger elongation usually leads to more complex shapes.
The rotational transform is a measure of the pitch of magnetic field lines, as it counts the number of poloidal transits of the field line as it completes a complete toroidal circuit $\iota=d\theta/d\phi=\mathbf B \cdot \nabla \theta / \mathbf B \cdot \nabla \phi$, and is a function of the magnetic flux $\psi$ only, $\iota=\iota'(\psi)$.
Both its value and its derivative, the magnetic shear $\iota'(\psi)$, are important for the confinement of fast particles as their radial orbit width can be estimated to be proportional to $1/\iota$ \cite{Figueiredo2024}.
Furthermore, rational flux surfaces, defined as locations where $\iota$ is a rational number, are associated with magnetic islands, which may be used as divertors in the edge of the device, but, when present in the core, may lead to loss of confinement.
The inverse aspect ratio $\epsilon=a/R$ is a measure of the effective minor radius $a$ of the device, given that the major radius $R$ in this work is fixed to 1.
Larger minor radii lead to a larger volume available to yield fusion energy, yielding therefore a large energy gain.
The magnetic well \cite{Mercier1964,Landreman2020a,Kim2021} is a measure of MHD stability, for which $V''(\psi)<0$, with $V(\psi)$ the volume enclosed by a flux surface of constant $\psi$, is defined as a necessary (but not sufficient) criterion for stability.

\begin{table}
\centering
{\footnotesize
\begin{tabular}{p{0.34\columnwidth} p{0.55\columnwidth}}
    \toprule
    \textbf{Column Name} & \textbf{Description} \\
    \midrule
    $\text{RBC}_{m,n}$ and $\text{ZBS}_{m,n}$ $\in$ $[-0.1, 0.1]$ & Fourier coefficients of the boundary. \\
    \midrule
    Quasisymmetry $\in$ $[0, 10]$ & Omnigenity of closed $B$ toroidal contours. \\
    \midrule
    Quasi-isodynamic $\in$ $[3.7 \times 10^{-6}, 0.6]$ & Omnigenity of closed $B$ poloidal contours.\\
    \midrule
    Rotational Transform $\in$ $[-2.4, 2.4]$ & Number of times a field line wraps poloidally for each toroidal circuit. \\
    \midrule
    Inverse Aspect Ratio $\in$ $[0.005, 0.1]$ & Ratio between minor and major radii. \\
    \midrule
    Magnetic Shear $\in$ $[-0.7, 0.5]$ & Radial derivative of the rotational transform. \\
    \midrule
    Magnetic Well $\in$ $[-0.5, 0.1]$ & Second radial derivative of the volume enclosed by a flux surface. \\
    \midrule
    Maximum Elongation $\in$ $[1.3, 309.9]$ & Ratio of semi-major and semi-minor axis of the cross-section. \\
    \midrule
    Mirror Ratio $\in$ $[0.009, 1.0]$ & Ratio of the maximum to the minimum $B$ along the magnetic axis. \\
    \midrule
    Number of Field Periods $n_{\text{fp}}=2$ & Discrete symmetry of the magnetic field. \\
    \midrule
    Convergence $\in$ $\{0, 1\}$ & Classification metric specifying if the numerical algorithm converged to the required tolerance of $\text{FTOL}<10^{-12}$. \\
    \bottomrule
\end{tabular}
}
\caption{Parameters stored in the stellarator database.
}
\label{tab:stellarator_columns}
\end{table}

The parameters used in this work are compiled in \cref{tab:stellarator_columns}.
A total of 21 columns are used.
The first eight columns contain the surface degrees of freedom $\mathbf x$ from \cref{eq:dofs}, while the remaining ones include the properties of the stellarator outlined before.
The last column of the database contains a convergence metric based on the ideal MHD solver, which indicates when VMEC converges without any error and holds values of either 0 or 1, indicating the absence or presence of convergence, respectively.
Due to the fact the Fourier coefficients are generated randomly, there are instances where VMEC is not able to yield a configuration either by encountering a surface with a Jacobian equal to zero, indicating that it intersects itself, or by the ideal MHD equation not having a sufficiently good tolerance, leading to a non-zero error flag.
In here, the tolerance is set by VMEC's input parameter $\text{FTOL}=10^{-12}$, which is higher when compared with published designs such as the ones in Ref. \cite{Landreman2022}, but it broadens the search and makes it more effective, as we are able to use only 16 radial grid points and a maximum of 1500 numerical iterations.
However, even for stellarators that converge, their high quasisymmetry or quasi-isodynamic values indicate poor performance, rendering them unhelpful for the training phase of the model.
Therefore, such stellarators are not included in the training data.
The database encompasses 12,421,004 configurations from which 457,488 have a quasisymmetry residual below 10. 

The database is constructed by sampling the Fourier coefficients of the plasma boundary, $\mathbf x$, using the pseudo-random generator from the $\text{numpy}$ library in Python \cite{Harris2020} based on an uniform distribution $p(x)=1/(b-a)$ with the limits $a$ and $b$ set to $-0.1$ and $0.1$, respectively.
While this limits the aspect ratio of the resulting stellarators, it provides for a more efficient search in parameter space.
The database is stored in Zenodo \cite{Laia2024} using a Structured Query Language format, SQL.
When predicting quasisymmetry and quasi-isodynamicity, the dataset is divided into features, $\mathbf{x}$, and targets, $\mathbf{y} =$ [quasisymmetry, quasi-isodynamicity].
As the values of $\mathbf y$ can span multiple orders of magnitude, a logarithm transformation is applied to $\mathbf y$ during training.

\section{Database of Stellarator Configurations}
\label{sec:database_properties}

\begin{figure}
    \centering
    \includegraphics[trim=0cm 2.0cm 0cm 0.0cm, clip, width=0.41\textwidth]{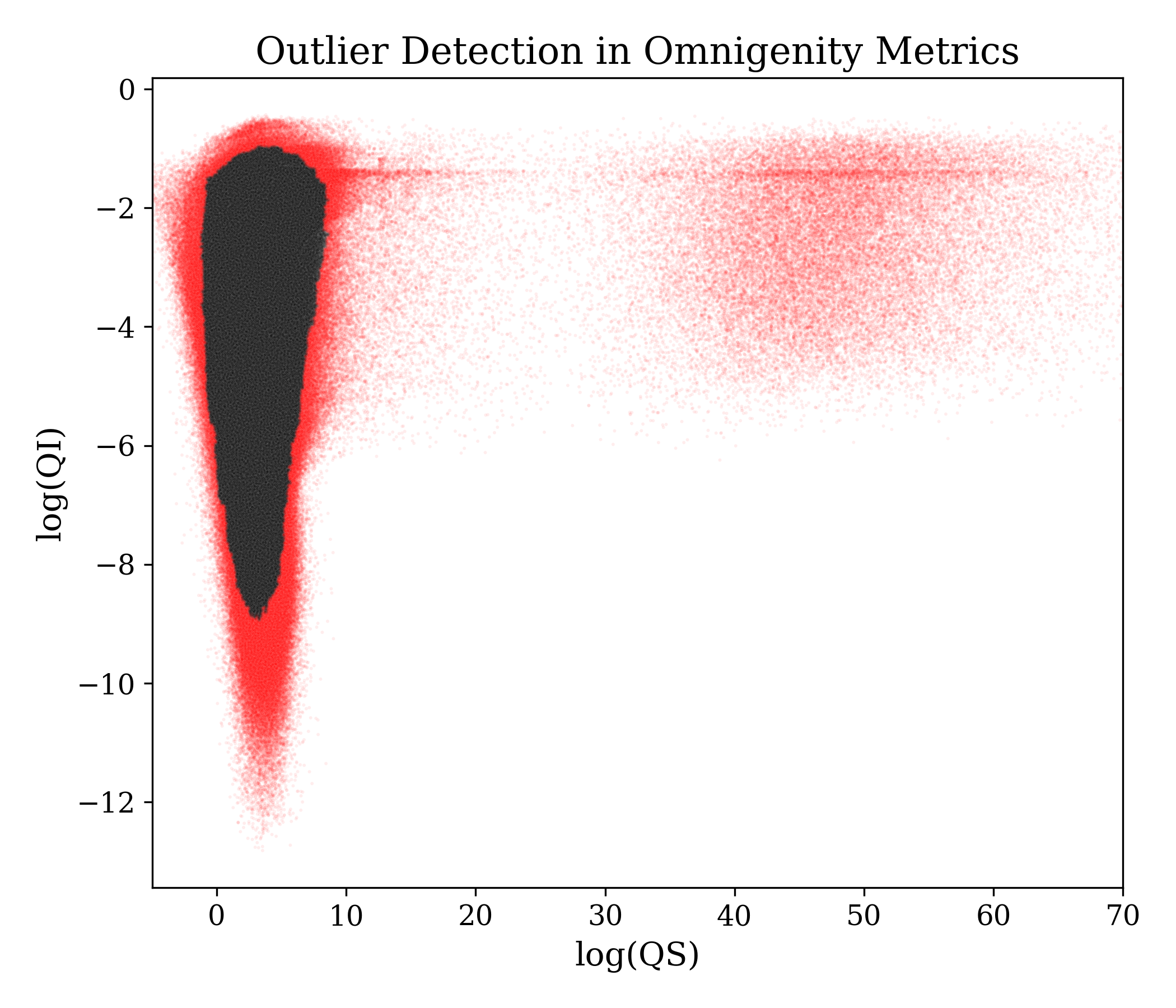}
    \caption{Red: anomalous datapoints as flagged by the Isolation Forest algorithm with a contamination rate of 10\%. Black: points kept when filtering for outliers.}
    \label{fig:outliers}
\end{figure}

We start the analysis of the resulting database by first applying an Isolation Forest Model \cite{Liu2008}, an unsupervised machine learning technique to detect outliers in the database and isolate anomalous datapoints.
In \cref{fig:outliers}, we show in red the anomalous datapoints as flagged by the Isolation Forest algorithm with a contamination rate of 10\% for the case of the log-transformed quasisymmetry and quasi-isodynamicity.
The remaining black points are used later when filtering for outliers.
In \cref{fig:outliers}, we see that values of quasisymmetry can span up to 100 different orders of magnitude, while quasi-isodynamicity spans up to 13.
Although the range of outliers seems to be large in the scatter plot, when applying the filtering, the percentage of data kept with both $\log(\text{QS})$ and $\log(\text{QI})$ filters is 92.21\%, while the percentage of data kept with $\log(\text{QS})$ and $\log(\text{QI})$ filters only is 95.12\% and 96.95\%, respectively.

\begin{figure*}
    \centering
    \includegraphics[trim=0cm 0.0cm 0cm 0.0cm, clip, width=0.99\textwidth]{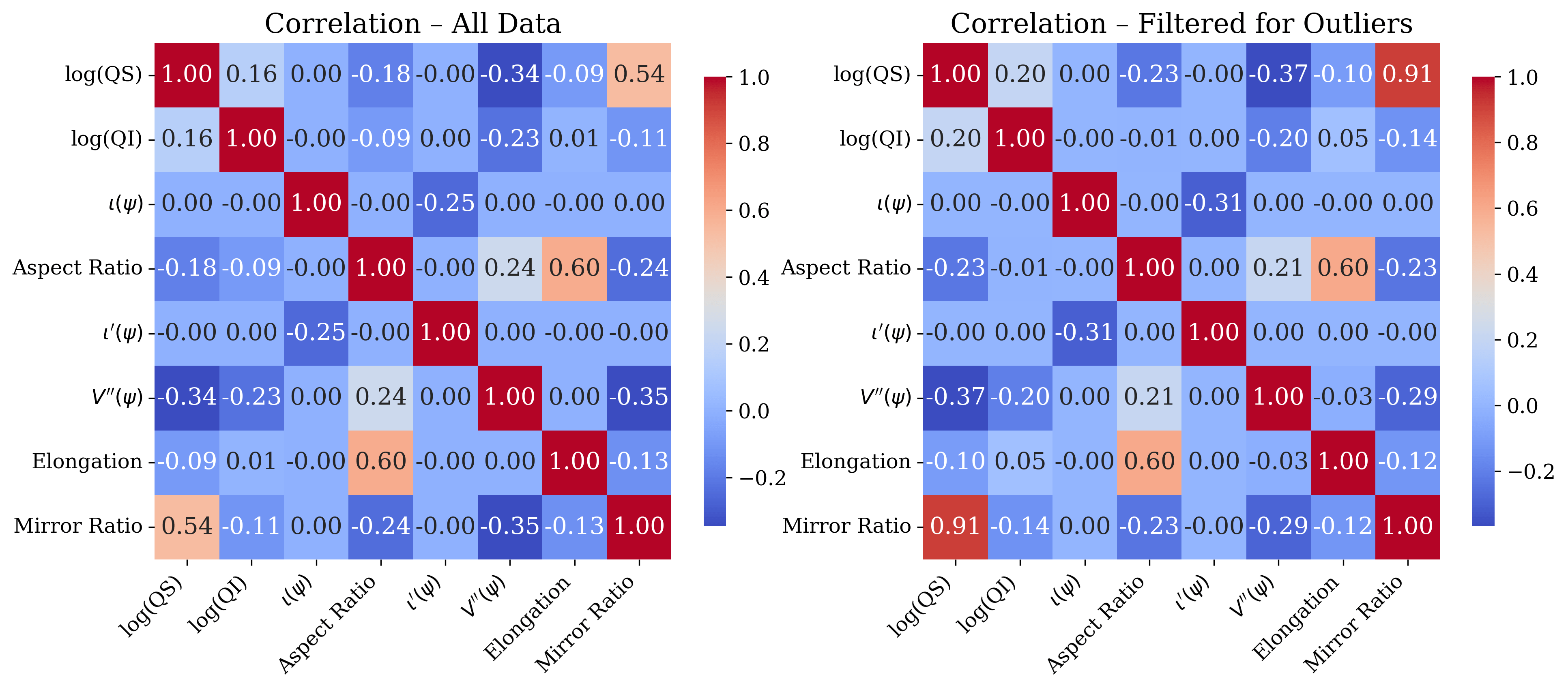}
    \caption{Correlation heatmap between physical metrics, both for every point in the database (left) and after filtering (right).}
    \label{fig:correlation}
\end{figure*}

In order to better understand the effect of the outliers and detect possible relations between physical metrics in the database, we show in \cref{fig:correlation} a correlation heatmap between every quantity present there.
The number in each cell is the pairwise Pearson correlation coefficient, indicating linear relationships between two features, with values ranging from $-1$ (perfect negative correlation) to $1$ (perfect positive correlation), and intermediate values indicating weaker or no linear dependence.
Positive correlations have a red color, while negative correlations have a blue color.
Two matrices are shown, with the correlations before and after the filtering for outliers is applied.
The two major correlations found are between quasisymmetry and mirror ratio, especially after the filtering, and between aspect ratio and elongation.
These are expected as quasisymmetry requires a constant magnetic field on-axis, which yields a mirror ratio of zero; therefore, for the larger aspect ratio stellarators studied here, the mirror ratio should be smaller with decreasing quasisymmetry residual.
Furthermore, as aspect ratio is a measure of the minor radius at locations of constant cylindrical angle, higher elongations may lead to a view of the stellarator at this angle that has higher aspect ratios.
We also mention the negative correlation of both QS and QI with positive magnetic well, as well as between rotational transform and shear.

\begin{figure}
    \centering
    \includegraphics[trim=0cm 0.0cm 0cm 0.0cm, clip, width=0.45\textwidth]{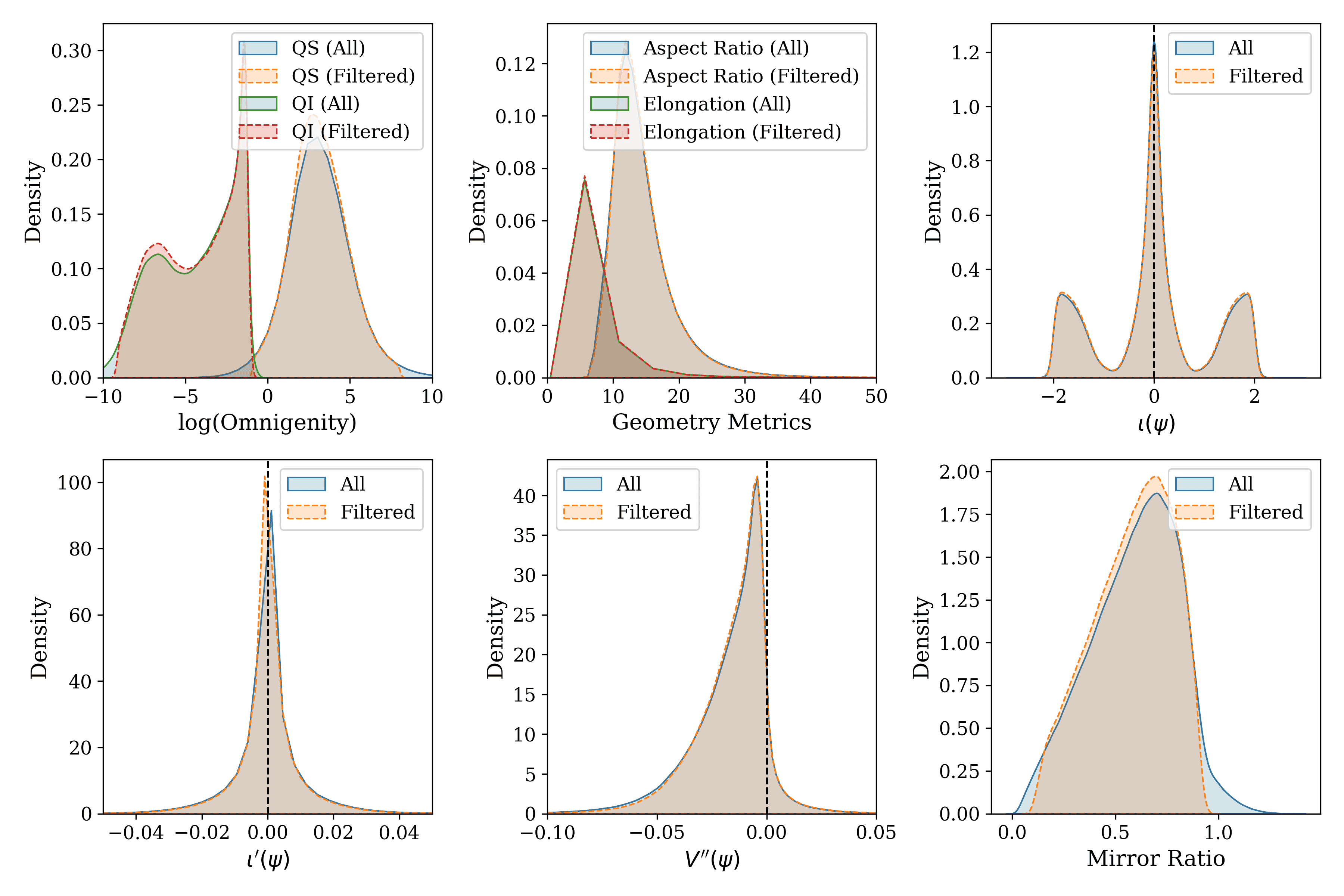}
    \includegraphics[trim=0cm 0.0cm 0cm 0.0cm, clip, width=0.45\textwidth]{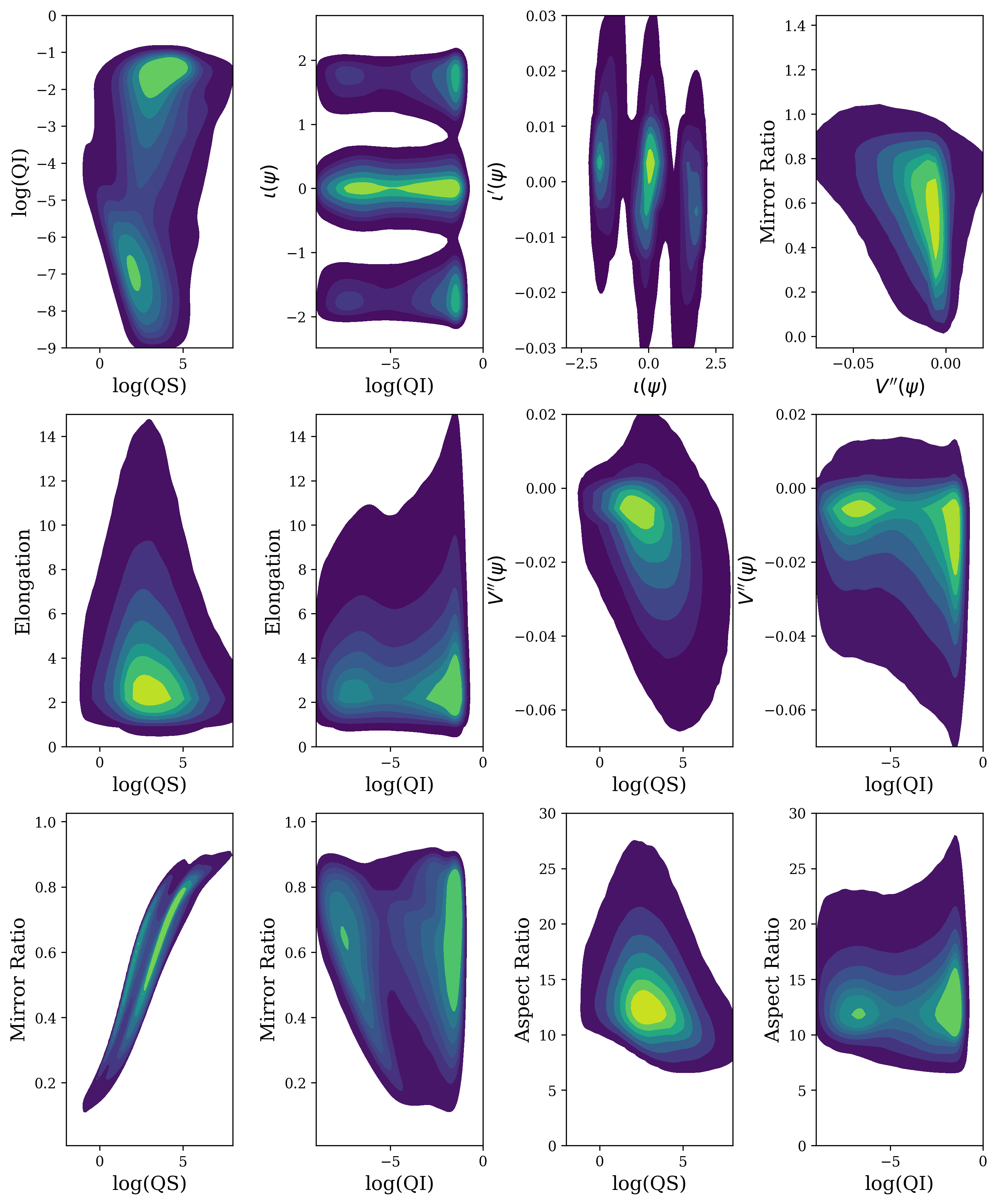}
    \caption{Top: Marginal distributions of QS, QI, and physics metrics using KDE smoothing. Bottom: two-dimensional KDE distributions.}
    \label{fig:distribution}
\end{figure}

In order to complement the information present in the correlation matrix, we show in \cref{fig:distribution} (top) the
distribution density of the different physics quantities.
Every curve in \cref{fig:distribution} (top) has unit area, and shows both the all data and the filtered data.
A Kernel Density Estimation (KDE) is applied with a Gaussian kernel to obtain a smoother marginal distribution.
We find that the removal of outliers has only a small effect on the distribution of QS and QI, as well as on the mirror ratio, where it removes mirror ratios very close to 1 and 0.
However, as seen in the correlation matrix, it makes the correlation between QS and the mirror ratio more striking.
The distribution shows that QS has a single peak, while QI has two peaks.
The rotational transform distribution is centered around zero.
However, there are configurations where the magnetic axis can possess non-zero helicity, leading to a peak at $\iota \sim 2$.
Finally, a bias towards negative values of $V''(\psi)$ is found, showing the difficulty of obtaining MHD stable configurations.

In order to complement the insights obtained with the marginal density distributions, we show in \cref{fig:distribution} (bottom) two-dimensional KDE estimation plots in order to highlight relationships between key physical metrics.
Density contours are shown at six levels to resolve the underlying structure of the distribution, and the data is filtered for outliers.
We find that quasisymmetry and mirror ratio have an approximately linear correlation at intermediate values of both quantities, and an inverse correlation is found for the QI case at smaller values of the QI metric.
While no such correlations are found for the other metrics, it is found that there are two large density locations of QS and QI, mostly with quasisymmetry values above 10.
This is consistent with the single KDE estimators where QI was seen to have two peaks.
In general, it is found that higher densities of the plots occur at larger values of QS and QI, and the peak of lower QI yields a magnetic hill, high mirror ratios, aspect ratios, and elongation.

\begin{figure*}[ht]
    \centering
    \includegraphics[trim=0cm 0.0cm 0cm 0.0cm, clip, width=0.9\textwidth]{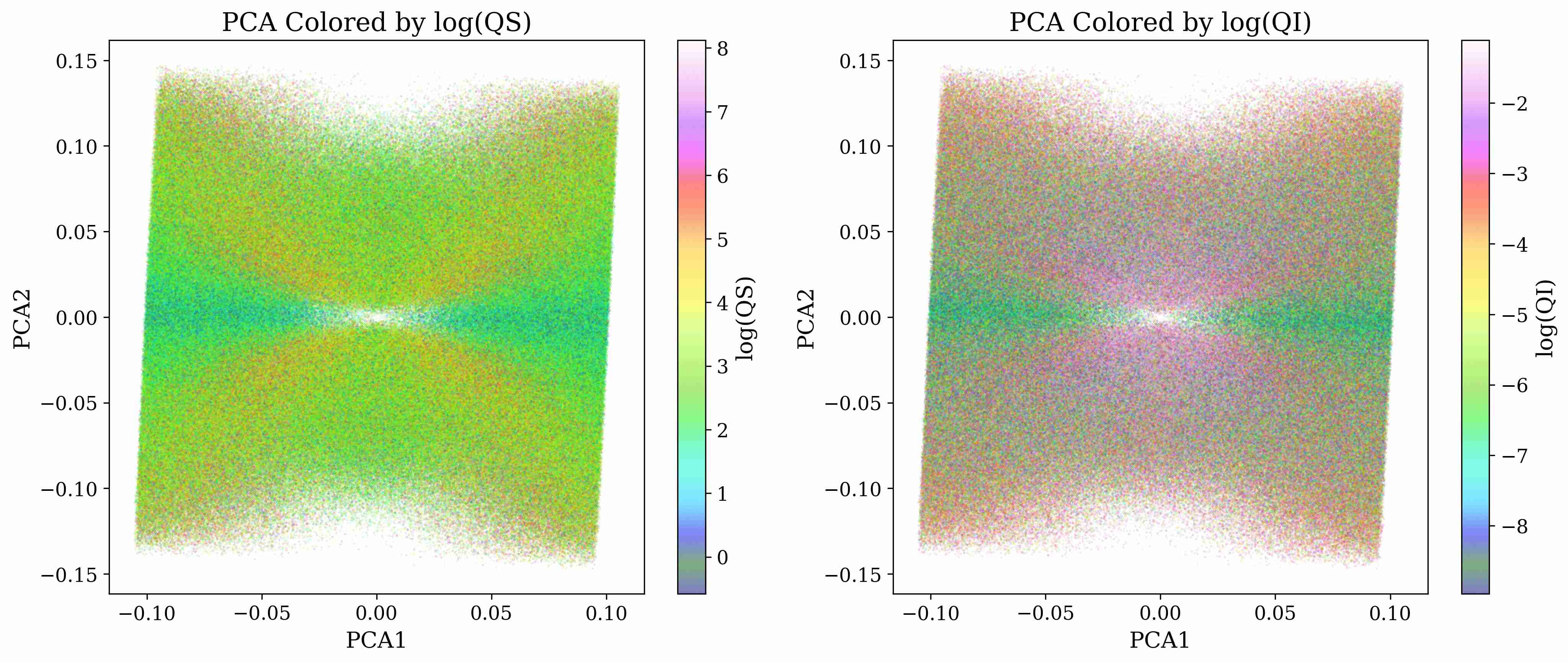}
    \caption{Dimensionality reduction of the surface Fourier coefficients using Principal Component Analysis PCA colored by the logarithm of the quasisymmetric (QS) and quasi-isodynamic metric (QI).}
    \label{fig:dimensionality_reduction_PCA}
\end{figure*}

We now analyze the database from the point of view of dimensionality reduction in order to decrease the large possible set of parameters defining the plasma boundary, in this case, Fourier coefficients $\text{RBC}_{m,n}$ and $\text{ZBS}_{m,n}$, to only two dimensions.
The first analysis involves a linear algorithm, Principal Component Analysis (PCA), and is shown in \cref{fig:dimensionality_reduction_PCA}.
This method seeks an orthonormal basis such that the projected variance of the data is maximized along the early directions.
We retain only the first two principal components and project the data into this reduced dimensionality representation.
In \cref{fig:dimensionality_reduction_PCA}, each point is colored by the logarithm of the QS (left) and QI (right) metric.
Such a visualization would allow us to assess if QS and QI have systematic variations along these two axes, if there are specific regions in Fourier space associated with improved omnigenity, or explore potential clusters for better optimization strategies.
We find that QS and QI residuals have regions of lower values along the axis of null PCA direction 2.
This motivates us to perform a nonlinear dimensionality reduction technique informed by the scalar values of QS and QI.

\begin{figure*}[ht]
    \centering
    \includegraphics[trim=0cm 0.0cm 0cm 0.0cm, clip, width=0.99\textwidth]{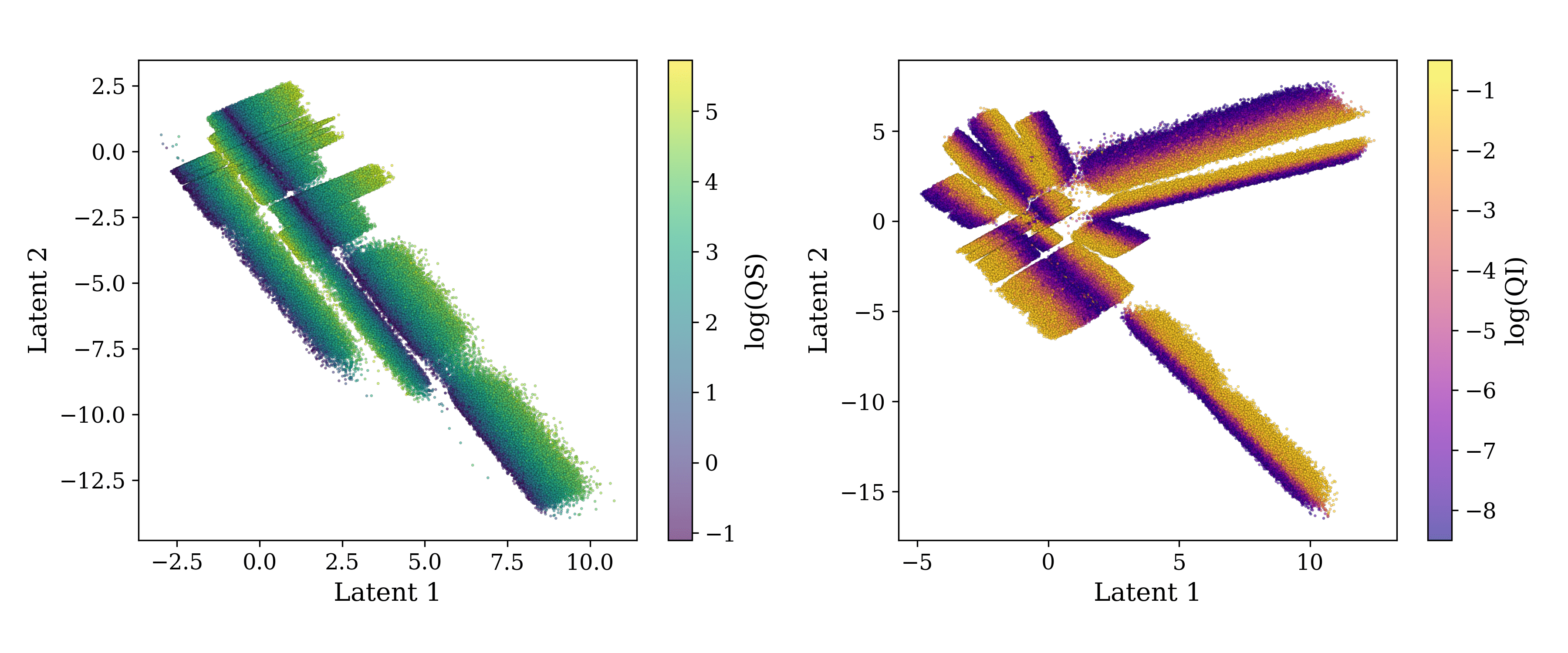}
    \includegraphics[trim=0cm 0.0cm 0cm 0.0cm, clip, width=0.99\textwidth]{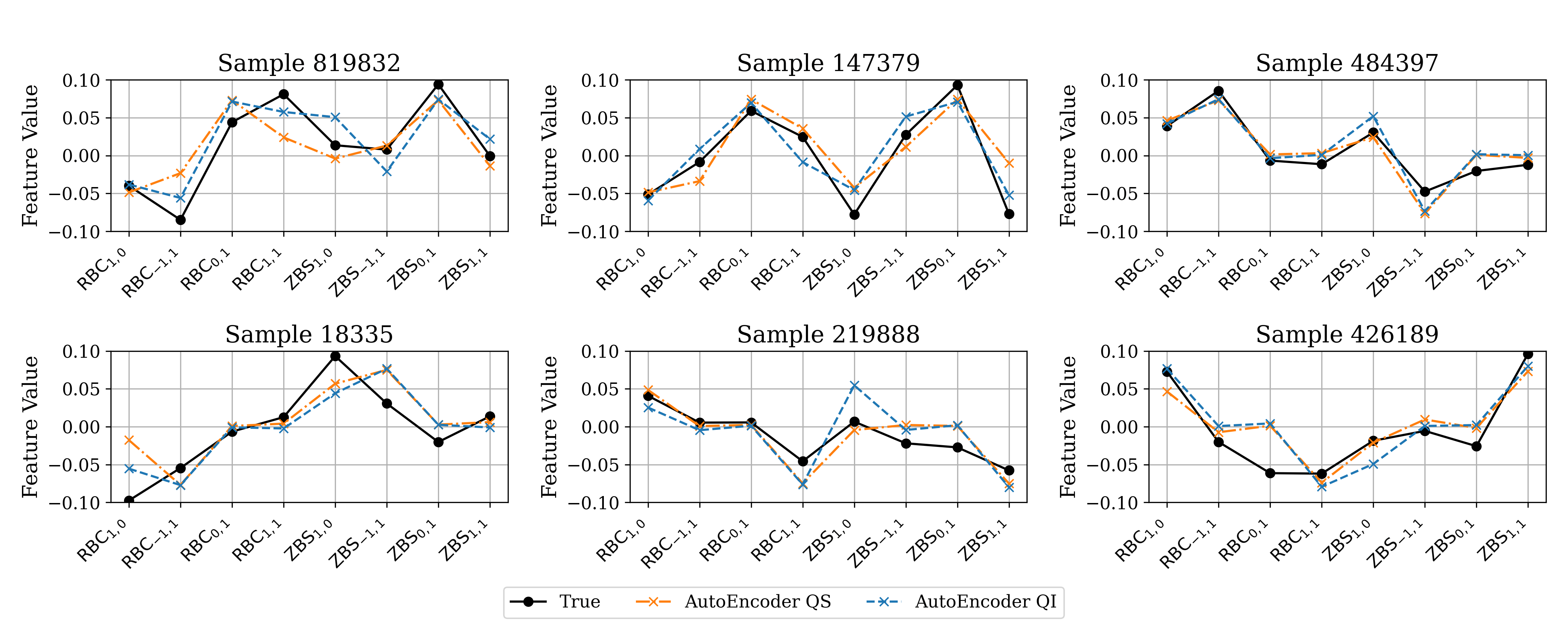}
    \caption{Supervised autoencoder model to reduce the dimensionality of the Fourier coefficients of the boundary and predict omnigenity.
    Top: latent space representation.
    Bottom: comparison between the true feature values and the reconstructed values for six random points in the database.
    }
    \label{fig:autoencoder}
\end{figure*}

\begin{figure*}[ht]
    \centering
    \includegraphics[trim=8cm 7.0cm 8cm 5.0cm, clip, width=0.55\textwidth]{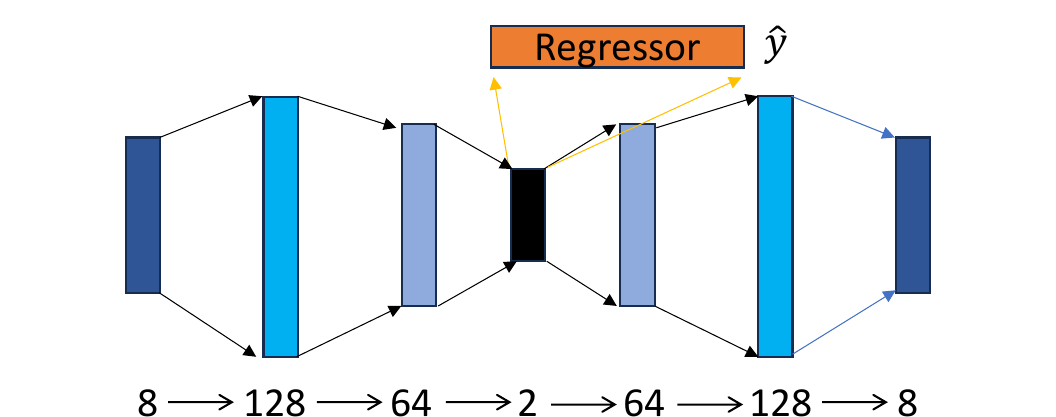}
    \caption{Encoder maps the eight boundary Fourier coefficients $x$ through two fully–connected layers (128 and 64 units, each followed by batch normalisation and ReLU) to a two–dimensional latent vector $z$. A regressor attached to $z$ predicts the confinement metric $\hat y\!\in\!\{\log f_{\mathrm{QS}},\,\log f_{\mathrm{QI}}\}$, while the decoder reconstructs $\hat x$.}
    \label{fig:autoencoder_achitecture}
\end{figure*}

For this purpose, we construct a supervised autoencoder approach \cite{Hinton2008}. The autoencoder architecture is shown in \cref{fig:autoencoder_achitecture}.
To ensure that the latent space captures directions that are genuinely relevant for omnigenity, we trained the auto-encoder in a supervised fashion. 
Inspired by the supervised-auto-encoder framework of Le, Patterson \&\ White~\cite{LePattersonWhite2018}, we couple the usual reconstruction loss with a regression head that predicts either \(\log f_{\mathrm{QS}}\) or \(\log f_{\mathrm{QI}}\) directly from the two-dimensional latent vector.  The total loss reads
\begin{equation}
  \mathcal{L}
  = (1-\lambda)\,\bigl\lVert \mathbf{x}-\hat{\mathbf{x}} \bigr\rVert_{2}^{2}
    + \lambda\,\bigl\lVert y-\hat{y} \bigr\rVert_{2}^{2},
  \label{eq:sae_loss}
\end{equation}
where \(\mathbf{x}\) is the vector of boundary Fourier coefficients, \(\hat{\mathbf{x}}\) its reconstruction, \(y\in\{\log f_{\mathrm{QS}},\log f_{\mathrm{QI}}\}\) is the target metric, and we fix \(\lambda=0.5\) to balance the two objectives.
The network, implemented in \textsc{PyTorch}, maps the eight Fourier coefficients \(\text{RBC}_{m,n}\) and \(\text{ZBS}_{m,n}\) into a two-dimensional latent space via a fully connected encoder with layer widths \(8 \!\to\! 128 \!\to\! 64 \!\to\! 2\). 
Each linear layer is followed by batch normalisation and a ReLU activation.  A symmetric decoder reconstructs the input, and a small regressor acting on the latent vector outputs the chosen confinement metric. 
We train for 1300 epochs with the Adam optimiser (initial learning rate \(10^{-3}\), decayed by \(2\times10^{-4}\) every 50 epochs) and a batch size of 512. 
All inputs and targets are standardised, and the training set is filtered exactly as in the preceding dimensionality-reduction experiments. 

The results of the supervised autoencoder are shown in \cref{fig:autoencoder}.
The resulting latent manifold (Fig.~\ref{fig:autoencoder}) exhibits a clear gradient along one axis that correlates with improved quasisymmetry or quasi-isodynamicity, while the orthogonal direction captures benign geometric variations.
In \cref{fig:autoencoder} (top), the latent space of the autoencoder is seen, where each point is a sample of the database in the reduced two-dimensional space, colored by its target metric $\log QI$ or $\log QS$.
Here, we find that the quasisymmetric model is able to find structure in the data, with most of the data lying in a narrow section of positive latent dimension 1 and negative latent dimension 2.
The quasi-isodynamic model shows more structure in the data with two clearly defined bands.
This is expected as it was found that $\log QI$ had two peaks in the distribution.
A comparison between the true feature values against the reconstructed values for 6 random points in the database is performed in \cref{fig:autoencoder} (bottom).
%
We find that the coefficient $\text{ZBS}_{1,0}$ is the one that is consistently harder to predict.
However, as seen in the next sections, this is also the parameter that contributes the least to the quasisymmetry and quasi-isodynamic metrics.
While such results are promising, the model used here should be more complex in order to more clearly capture the structure in the data and be able to reconstruct it with minimal error.
This can include the use of variational autoencoders or additional regularization techniques to reduce overfitting. 


\section{Gradient Models: LightGBM and LightGBM LSS}
\label{section:MLmodels}

We now introduce the machine learning models based on Gradient Boosting used to analyze and predict omnigenity in stellarators.
These are classification and regression models based on the LightGBM algorithm \cite{Ke2017, Ma2018}.
LightGBM is an open-source framework that can handle large datasets and efficiently produce highly accurate predictive models.
This framework uses the concept of gradient boosting, which involves training new models to correct the errors made by previous ones, iteratively improving the predictive performance \cite{Natekin2013}.
Its code is openly available in \cite{lightgbm_gitrepo}.
Besides machine learning tasks, it has recently been extended to physics-constrained algorithms \cite{Chai2022}.

We first detail the regression models used and later the classification model.
A regression model is a predictive model that estimates the relationship between a dependent variable and one or more independent variables, in this case, quasisymmetry and quasi-isodynamicity.
LightGBM uses decision trees \cite{Song2015}, where decisions are made at various internal nodes based on feature values.
The final values are assigned at the leaf nodes that represent points where the data is split based on certain conditions, and leaves represent the final output of the tree.
A unique feature of LightGBM is its use of leaf-wise growth for building decision trees.
%
%
Comparing the traditional approach, known as depth-wise growth, to leaf-wise growth, the former splits all leaves at the same depth level before moving on to the next level.
This creates more balanced trees, which are generally simpler and less prone to overfitting.
On the other hand, in leaf-wise growth, the algorithm chooses the leaf with the highest potential for reducing error and splits it, resulting in deeper and narrower trees.
These tend to require fewer splits to achieve the same accuracy as depth-wise growth.


The flexibility of LightGBM can be enhanced by modeling and predicting the full conditional distribution of a univariate target as a function of covariates.
This is performed using the probabilistic version of LightGBM, LightGBMLSS \cite{Marz2022}, which offers more than point predictions by capturing the uncertainty in the model's outputs.
Therefore, we can understand where the model is more certain of its predictions. 
While LightGBM focuses on predicting a single outcome or classification from a set of input features, LSS extends this by modeling the entire probability distribution of the target variable.

LightGBM LSS selects the most probable point from the predicted distribution as the expected point.
It interprets the loss function from a statistical perspective by using Maximum Likelihood estimation, meaning that instead of minimizing the error between predicted and actual values, as traditional LightGBM, it estimates the parameters of the probability distribution that best fits the data.
The process involves multi-parameter optimization, i.e., simultaneously optimizing multiple parameters that describe the distribution of the response variable.
To achieve this, LightGBM LSS grows separate decision trees for each of these distributional parameters.
Each tree is specifically trained to predict one parameter of the distribution.

For the LightGBM LSS algorithm, determining the most suitable distribution for the output values was essential for enhancing the model's predictive accuracy.
To accomplish this, multiple candidate distributions were tested.
The main goal was to identify which distribution most accurately represents the quasisymmetry and quasi-isodynamic data, and the methodology involved using the LightGBMLSS library, which provides a range of distribution models and utilities for evaluating their suitability.
The first part involved evaluating a variety of individual distributions to determine their fit to the data, such as Gaussian, StudentT, Gamma, Cauchy, LogNormal, Weibull, Gumbel, Laplace, Beta, Poisson, SplineFlow, and NegativeBinomial.
In the second part of the analysis, a more complex approach was taken by evaluating mixtures of Gaussian distributions.
The candidate distributions, in this case, were mixtures with varying numbers of components $M$, from two to nine.
Furthermore, a Gumbel-Softmax distribution is added \cite{Jang2017} with a non-negative scalar temperature $\tau$ that controls the sharpness of the output distribution.
As $\tau$ approaches 0, the mixing probabilities become more discrete, and as $\tau$ increases, the mixing probabilities become more uniform.
Gumbel-softmax 
The two arguments of the mixture function that best suit the quasisymmetry distribution were found to be $M=9$ and $\tau=1.0$.
Regarding the quasi-isodynamic distribution, the parameters found were $M = 2$ and $\tau = 1.0$.
Knowing the distribution, the next step is to find the best hyperparameters for the model and train it.

Machine learning models such as LightGBM rely on a set of parameters, known as hyperparameters, to control the learning process.
Selecting the best hyperparameters is crucial as they influence the model's complexity, behavior, speed, and overall performance.
These parameters must be chosen carefully to minimize the selected loss function.
All models had the mean squared error (MSE) as their loss function, including the neural networks in \cref{section:FFNN}, measuring how well the model's predictions match the actual outcomes.
However, manually selecting these values is time-consuming and prone to bias.
To address these challenges, the Optuna \cite{Akiba2019} framework is employed.
Optuna is an open-source hyperparameter optimization framework designed to automate and efficiently manage the tuning of hyperparameters for machine learning models.
It formulates hyperparameter optimization as the process of minimizing or maximizing an objective function that takes hyperparameters as input and returns a validation score. 

We now describe the hyperparameter optimization process used with Optuna and LightGBM, where the following parameters were varied: boosting type, maximum depth, number of leaves, minimum data, learning rate, data sampling strategy, linear tree option, tree learning algorithm, and distributed learning algorithm.
The first parameter to be optimized is the boosting type, either Gradient Boosting Decision Tree (\textit{GBDT}) \cite{Marz2022}, Dropouts meet Multiple Additive Regression Trees (\textit{DART}) \cite{Rashmi2015}, and Random Forest (\textit{RF}) \cite{Zhang2012}.
The maximum depth controls how deep the tree can grow, while the number of leaves determines the maximum number of leaves in a tree, influencing the model's complexity.
The minimum data in a leaf specifies the minimum number of data points allowed in a leaf, with higher values enforcing simpler models and lower values allowing for more complex models.
The learning rate influences the step size at each iteration, guiding the model toward minimizing the loss function.
The data sampling strategy was chosen between \textit{bagging} and \textit{GOSS} (Gradient-based One-Side Sampling) \cite{Ke2017}, with \textit{GOSS} improving efficiency by focusing on the most informative samples based on their gradient values.
The linear tree option \cite{Shi2019} enables the integration of linear models at the leaves, capturing linear relationships within the data more effectively.
The tree learner algorithm was chosen from \textit{voting}, \textit{data}, \textit{feature}, and \textit{serial}.

The distributed learning algorithm can be chosen from feature parallel or data parallel algorithms.
In the feature parallel algorithm, each machine works on a different set of features, finds the local best split, communicates with other machines to determine the overall best split, and then performs the split.
While this approach reduces some overhead, it still requires significant communication to share split results, which can be costly for large datasets.
Data parallelism, on the other hand, partitions the data horizontally across different machines.
Each machine uses its local data to construct histograms for all features, and then these local histograms are merged to find the global best split, although such a method incurs high communication costs due to the need to merge histograms.
Finally, voting in parallel further reduces costs by implementing a two-stage voting process for feature histograms.
Instead of relying solely on global histogram information, this method leverages local statistical information from each machine, significantly minimizing data transfer during the split and balancing communication efficiency \cite{Meng2016}.

The hyperparameter optimization for the LightGBM LSS case is performed using the built-in \textit{hyper\_opt} function \cite{Marz2022}.
This function also leverages Optuna for the optimization process.
The boosting type, maximum depth, number of leaves, and minimum data parameters are similar to LightGBMLSS, while LightGBMLSS further considers the hyperparameters minimum gain, minimum sum of Hessians, subsample ratio, subsample frequency, feature fraction, and feature pre-filter.

The minimum gain to split helps in pruning the tree by preventing splits that do not provide a significant gain in reducing the loss function.
The minimum sum of Hessians in a leaf parameter determines the minimum sum of Hessian values (a measure related to the second derivative of the loss function) required in a leaf, ensuring that the leaves are robust and not influenced by small fluctuations in the data.
The subsample ratio defines the fraction of data to be used for training each tree, and the subsample frequency determines how often subsampling is used.
The feature fraction selects a fraction of features used for building each tree.
The feature pre-filter was set to false, ensuring that all features were considered during training. 

The specific ranges for the LightGBM and LightGBM LSS numerical hyperparameters are provided in \cref{tab:hyperparameters}.
The optimization results and the model performance is analyzed in the next section.

\begin{table}[h!]
    \centering
    \footnotesize
    \begin{tabular}{l c c}
        \toprule
        \textbf{Hyperparameter} & \textbf{LightGBM} & \textbf{LightGBM LSS} \\
        \midrule
        Maximum Depth & 1 to 150 & 1 to 40 \\
        Number of Leaves & 2 to 5000 & 2 to 500 \\
        Minimum Data \\in a Leaf & 50 to 7000 & 20 to 2000 \\
        Learning Rate & 0.01 to 0.5 & 0.01 to 0.4 \\
        Minimum Gain \\to Split & - & 0.01 to 40 \\
        Minimum Sum \\Hessian in a Leaf & - & 0.01 to 100 \\
        Subsample Ratio & - & 0.5 to 1.0 \\
        Subsample Frequency & - & 1 to 20 \\
        Feature Fraction & - & 0.3 to 1.0 \\
        \bottomrule
    \end{tabular}
    \caption{Range of hyperparameters for LightGBM and LightGBM LSS optimization.}
    \label{tab:hyperparameters}
\end{table}

\section{Gradient Boosting Model Results}
\label{sec:gbm}

In this section, we detail the gradient boosting model parameters, including the LightGBM and LightGBM LSS versions, for both the prediction of quasisymmetry and quasi-isodynamicity.
For quasisymmetry, the LightGBM model achieved a mean squared error of 0.296.
The optimal hyperparameters are found to be the following: boosting type \textit{dart}, a max depth of 112, 1665 leaves, a minimum data in leaf value of 50, a learning rate of 0.1921, 2853 iterations, a data sample strategy of \textit{bagging}, and a tree learner set to \textit{feature}.
After tuning with LightGBM LSS, the best hyperparameters were a max depth of 12, 28 leaves, a minimum data in leaf value of 1301, a minimum gain to split of 0.4320, and a minimum sum hessian in leaf of 0.3016. The sub-sample ratio was 0.8580, with a sub-sample frequency of 17, and a feature fraction of 0.9625. The learning rate was adjusted to 0.3294, with the feature pre-filter set to False. The boosting method remained \textit{DART}, and 200 optimization rounds were used.

For quasi-isodynamicity, the LightGBM model achieved a mean squared error of 1.246.
The optimal hyperparameters included boosting type \textit{dart}, a max depth of 139, 1684 leaves, a minimum data in leaf value of 51, a feature fraction of 0.9631, a learning rate of 0.1265, 2707 iterations, and a tree learner set to \textit{feature}.
After LightGBM LSS tuning, the best hyperparameters were a max depth of 113, 114 leaves, a minimum data in leaf value of 1537, a minimum gain to split of 0.0117, and a minimum sum hessian in leaf of 0.0132. The feature fraction was increased to 0.9786, with a learning rate of 0.3458, and the feature pre-filter was set to False. The boosting method was switched to \textit{GBDT}, and 200 optimization rounds were employed.

After training the LightGBM LSS models, we can now perform predictions of quasisymmetry and quasi-isodynamicity in stellarators.
For this purpose, 15,000 random samples are drawn from the predicted distributions.
These samples represent multiple possible outcomes based on the predicted distribution for each data point.
Subsequently, actual and predicted density functions are assessed by comparing the real distribution of the target variable and the distributions predicted by the model.
Finally, SHAP (SHapley Additive exPlanations) \cite{Lundberg2017APredictions} graphs are used to find the discriminative power between $\mathbf{X}$'s components, quasisymmetry, and quasi-isodynamicity.

\begin{figure}
\centering
\includegraphics[trim=0.0cm 0.2cm 0.2cm 1.1cm, clip, width=0.95\linewidth]{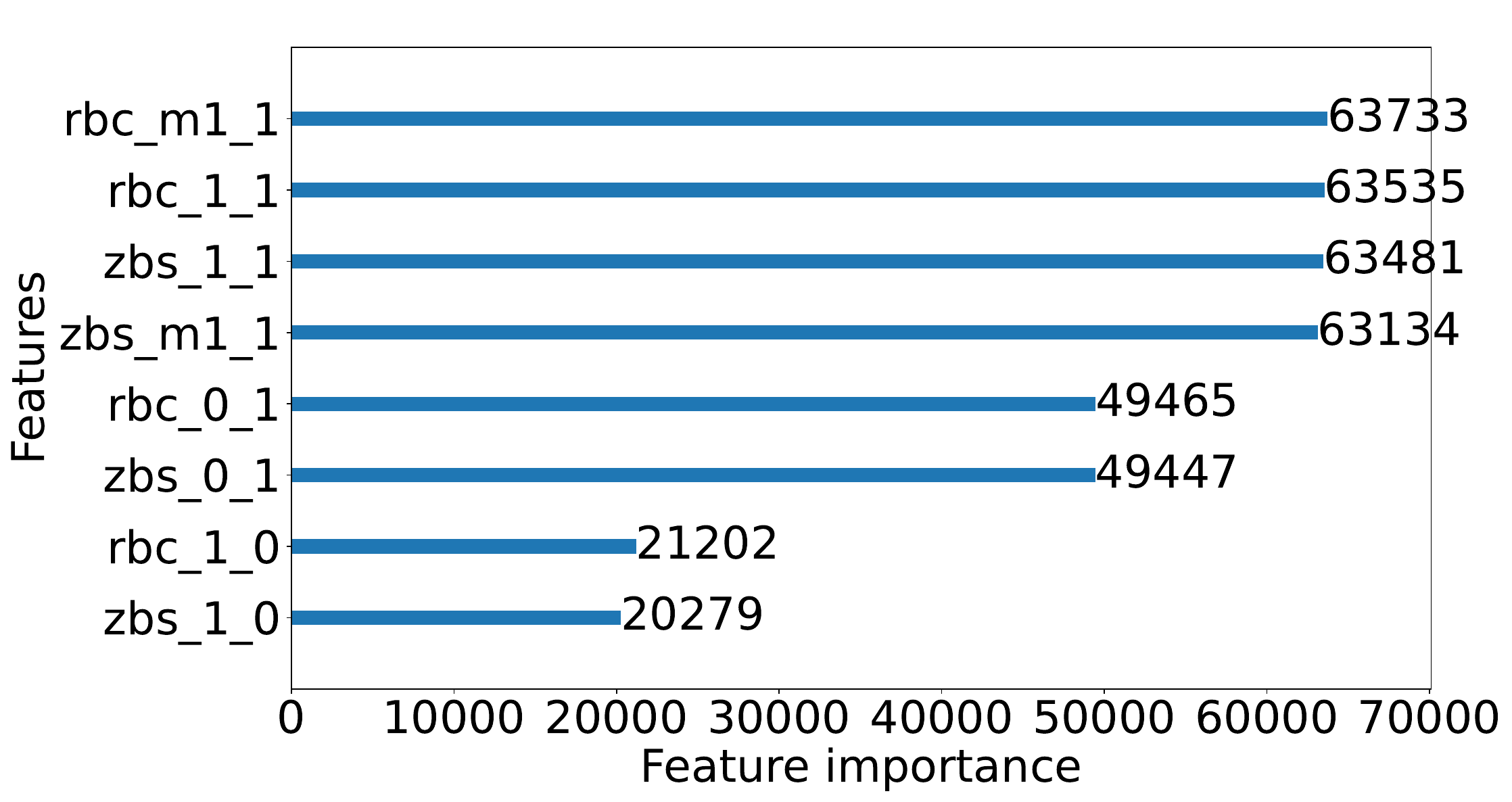}
\caption{
Feature importance
for the classification model to predict if VMEC is able to converge.}
\label{fig:feature_matrix}
\end{figure}

We start by analyzing the performance of both classification and regression models trained using LightGBM and LightGBM LSS.
For the classification task, we evaluated the model's ability to predict if VMEC can converge and whether quasisymmetry has a value below 10, using metrics such as accuracy, precision, and recall.
We also computed the resulting normalized confusion matrix to provide a detailed breakdown of the model's predictions. 
In \cref{tab:classification_results}, label 1 represents the cases where VMEC ran without error and quasisymmetry has a residual value below 10, and label 0 represents the cases where these conditions were not verified.
The metrics used for each label are precision and recall, which can help to understand the false positives and the false negative predictions.
For label 1, precision measures the proportion of predicted positive instances that are positive.
High precision means there are very few instances where the model predicted a quasisymmetry below 10, but the stellarator did not have this characteristic or did not converge at all.
Recall represents the proportion of actual positive instances that are correctly predicted.
High recall means there are very few cases where stellarators that converged and had a quasisymmetry below the selected threshold were incorrectly predicted as not having these characteristics.
Notably, the model presented better results in identifying stellarators that fail these conditions.
The classification model trained using LightGBM effectively predicted VMEC convergence and quasisymmetry below 10, predicting 99.7\% of the configurations with those conditions correctly.
Additionally, 91,500 devices from the test set converged, and the model accurately identified 77,848, correctly predicting 85.1\% cases of this label.
%
%
%
We show the feature importance on the prediction of VMEC convergence in \cref{fig:feature_matrix}
.
This is performed natively with LightGBM as it quantifies the total number of times a feature is used to split the data across all decision nodes in the ensemble of boosted trees. 
The features that most significantly influenced the output were identified as $\text{RBC}_{1,1}$, $\text{RBC}_{-1,1}$, $\text{ZBS}_{-1,1}$, and $\text{ZBS}_{1,1}$. 
Conversely, the features that least influenced the convergence were $\text{RBC}_{1,0}$ and $\text{ZBS}_{1,0}$.

\begin{table}
\centering
\footnotesize
\begin{tabular}{c c c c}
\toprule
\textbf{} & \textbf{Precision} & \textbf{Recall} & \textbf{Number of instances} \\ 
\midrule
0 & 0.997 & 1.00 & 2,399,336 \\ 
1 & 0.92 & 0.85 & 91,500 \\ 
\bottomrule
\end{tabular}
\caption{Table containing metrics that enable the evaluation of the classification model, i.e., precision, recall, and total number of observations for each label in the test set.}
\label{tab:classification_results}
\end{table}

\begin{figure}[ht]
\centering
\includegraphics[trim=0.3cm 0.0cm 0.2cm 0.7cm, clip, width=0.48\textwidth]{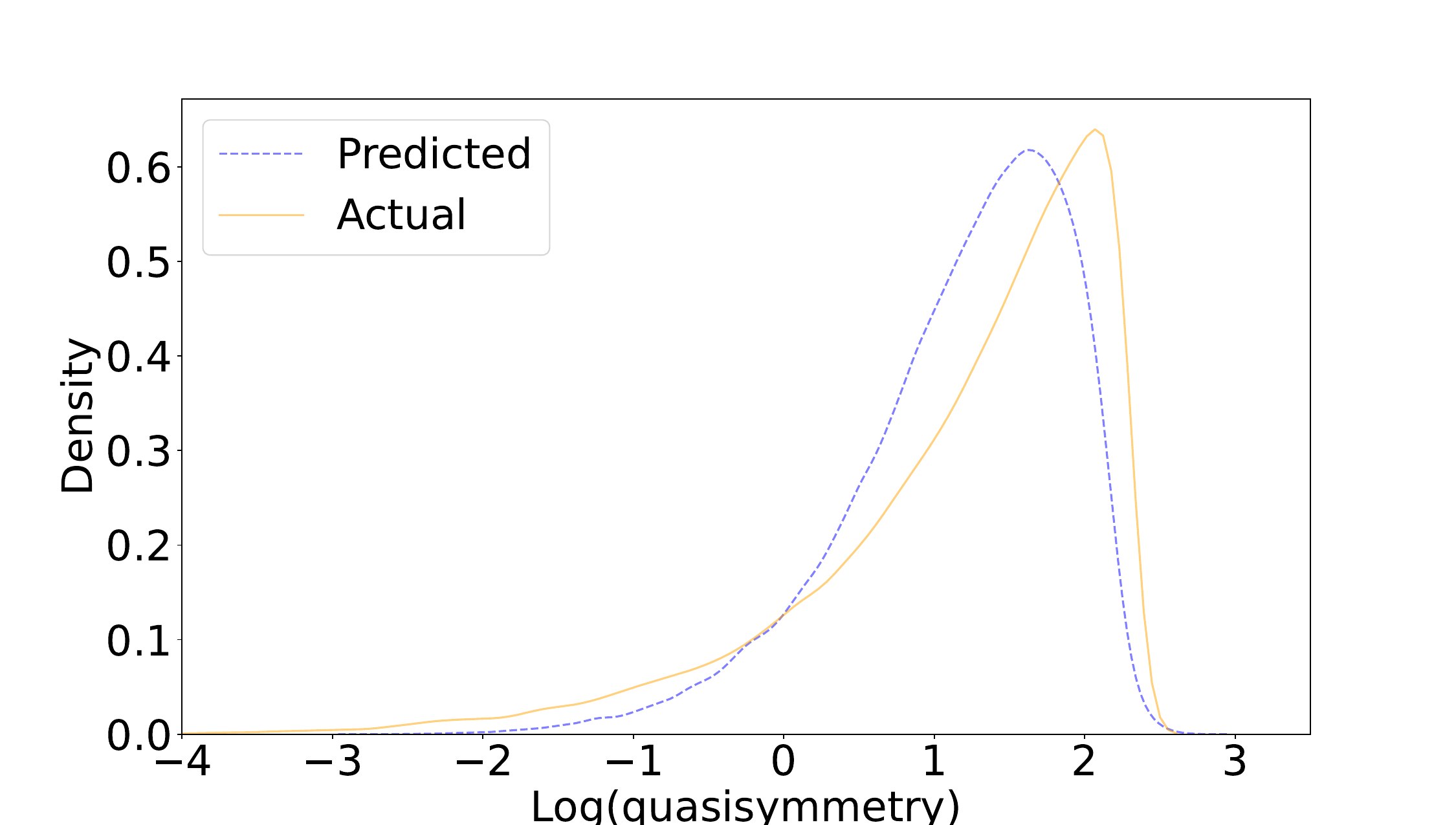}
\includegraphics[trim=0.3cm 0.3cm 0.2cm 0.7cm, clip, width=0.42\textwidth]{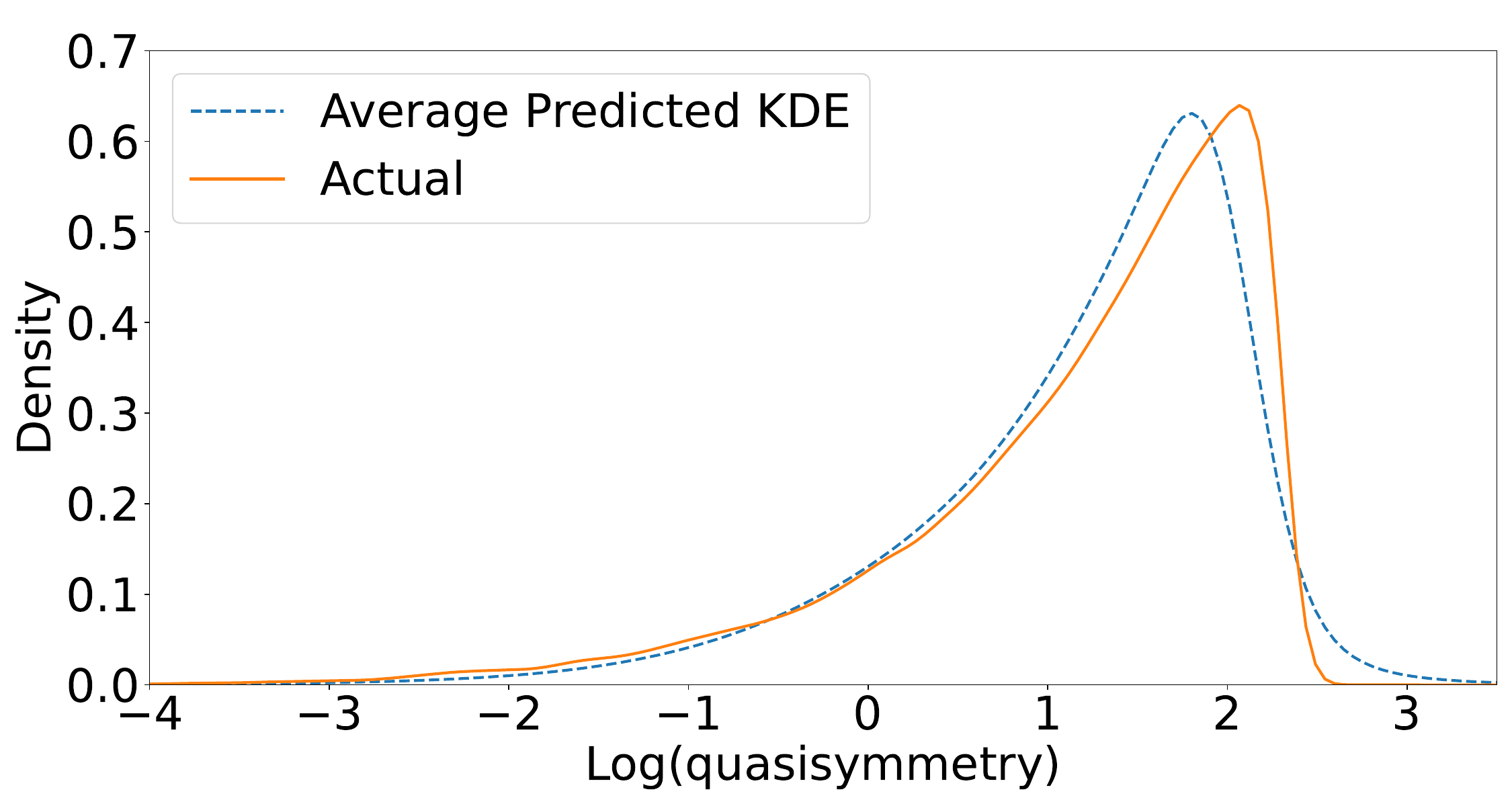}
\caption{Comparison of actual and predicted distributions using LightGBM (top) and LightGBM LSS (bottom). Both plots illustrate the log(quasisymmetry) on the X-axis and the corresponding predicted density on the Y-axis.}
\label{fig:actual_vs_pred_qs}
\end{figure}

\begin{figure}[ht]
\centering
\includegraphics[trim=0.3cm 0.3cm 0.2cm 0.7cm, clip, width=0.41\textwidth]{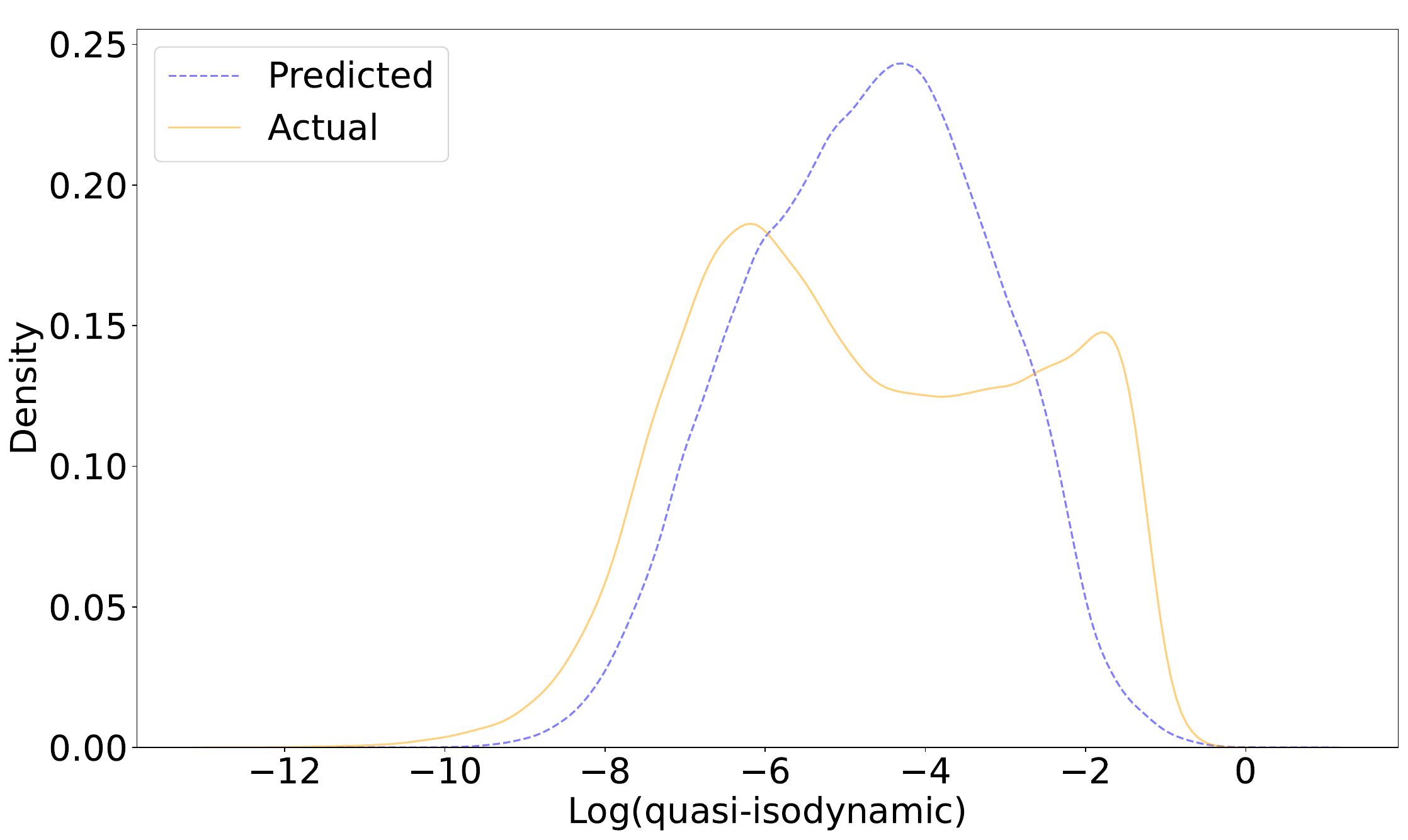}
\includegraphics[trim=0.3cm 0.3cm 2.9cm 2.1cm, clip, width=0.41\textwidth]{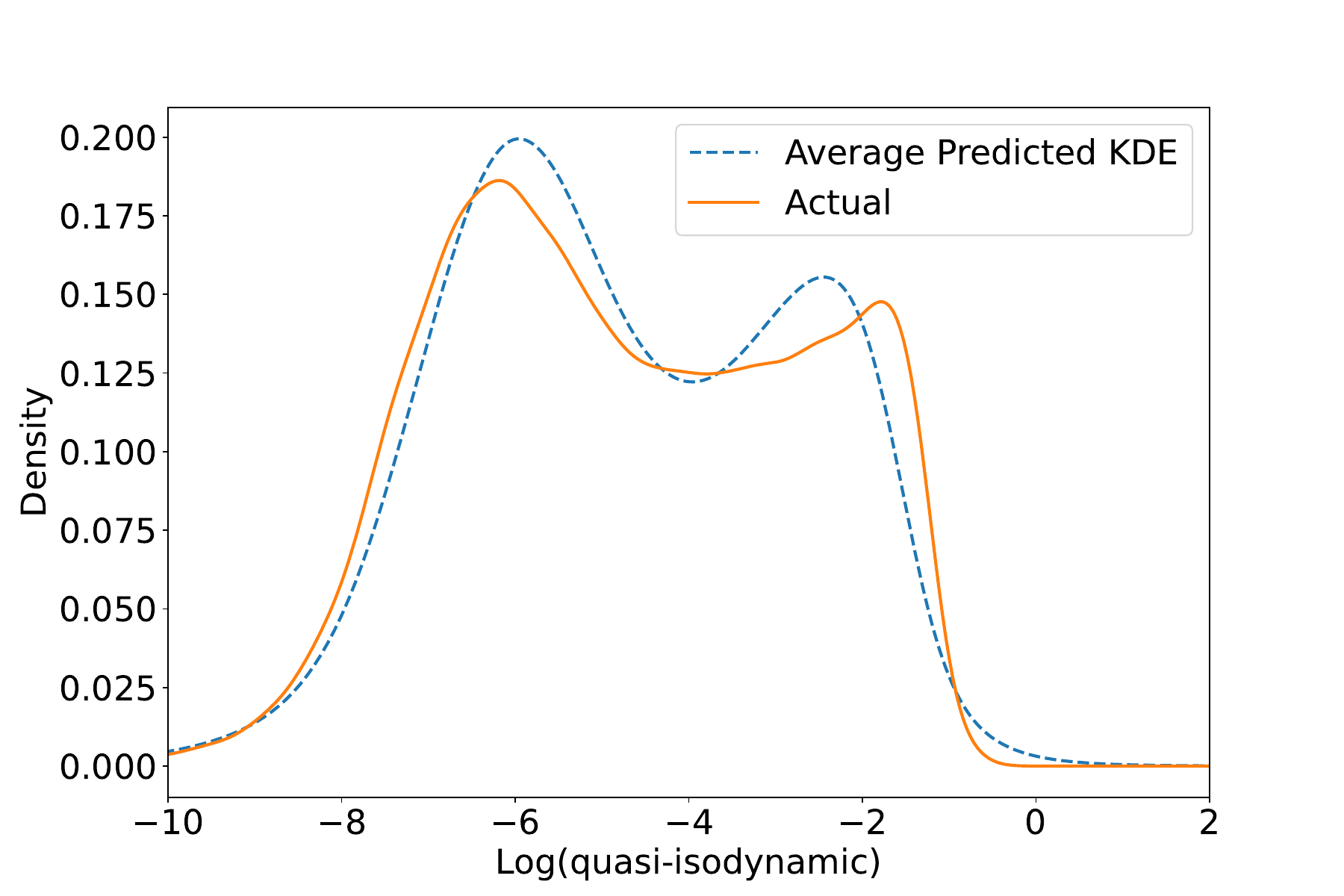}
\caption{Comparison of actual and predicted distributions using LightGBM (top) and LightGBM LSS (bottom). Both plots illustrate the log(quasi-isodynamic) on the X-axis and the corresponding predicted density on the Y-axis.}
\label{fig:actual_vs_pred_qi}
\end{figure}

\begin{figure}[ht]
\centering
\includegraphics[trim=3.1cm 0.0cm 4.5cm 3.0cm, clip, width=0.46\textwidth]{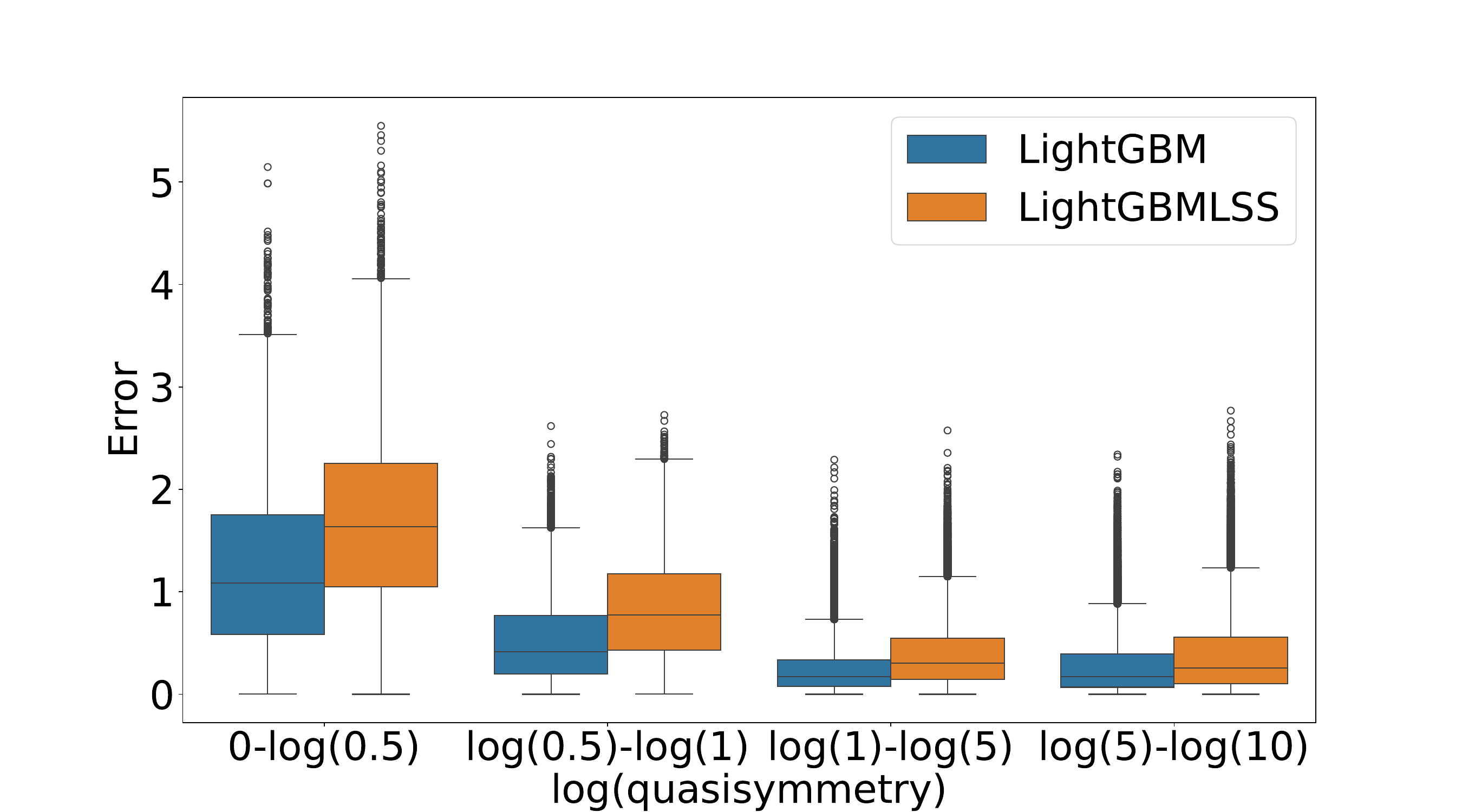}
\includegraphics[trim=3.1cm 0.0cm 4.5cm 3.0cm, clip, width=0.46\textwidth]{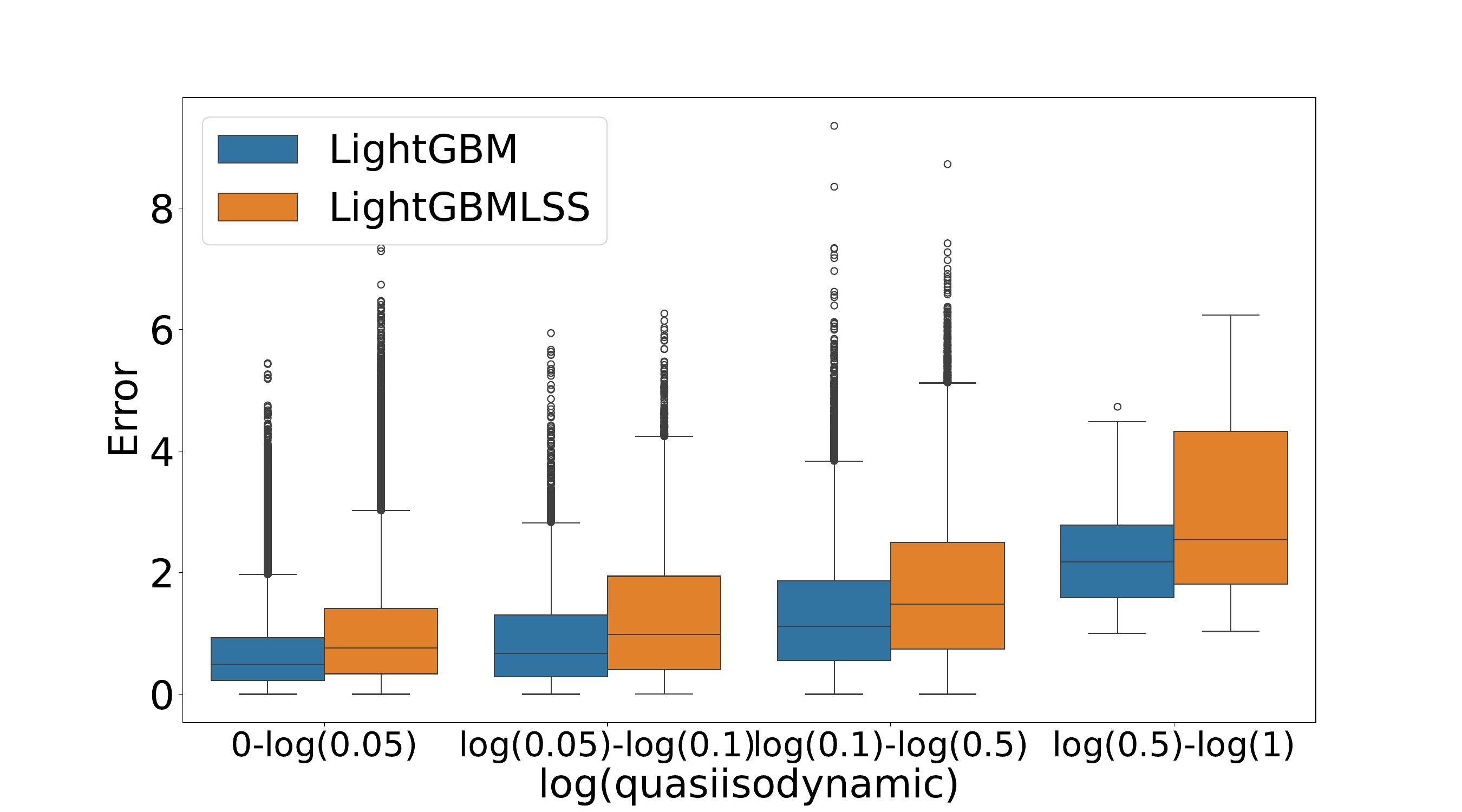}
\caption{Boxplot with MSE errors for both models, LightGBM (blue) and LightGBM LSS (orange), for quasisymmetry (top) and the quasi-isodynamic distributions (bottom).}
\label{boxplot}
\end{figure}

The resulting distributions of the LightGBM (top) and LightGBM LSS (bottom) regressor models are shown in \cref{fig:actual_vs_pred_qs} for the quasisymmetric model and \cref{fig:actual_vs_pred_qi} for the quasi-isodynamic model.
In both cases, the predicted distribution using the probabilistic model is illustrated by the mean of the kernel density estimate plots derived from 2,000 out of the 15,000 drawn samples.
It is seen that, in general, the probabilistic LSS model is able to obtain marginal distributions that are closer to the actual distribution.
This is especially true for the two-peaked distribution of QI stellarators.
However, while the distribution is closer in the probabilistic model than the non-probabilistic one, as shown in \cref{boxplot} (top), evaluating the boxplot of errors revealed that the LightGBM model generally had lower prediction errors compared to the LightGBM LSS model across most quasisymmetry values.
This suggests that LightGBM is more suitable for this specific dataset and prediction task, especially considering its fewer and smaller outliers compared to LightGBM LSS.
Additionally, the errors were larger at lower quasisymmetry values, likely due to the scarcity of low quasisymmetry values in the database.
On the other hand, \cref{boxplot} (bottom) illustrates that the errors for LightGBM LSS were more pronounced across the logarithm of the quasi-isodynamic distribution, particularly for higher values. This increase in error might be related to the reduced amount of data available as quasi-isodynamicity increases. 

\begin{figure}[ht]
\centering
\includegraphics[trim=0.3cm 0.3cm 0.7cm 1.1cm, clip, width=0.5\textwidth]{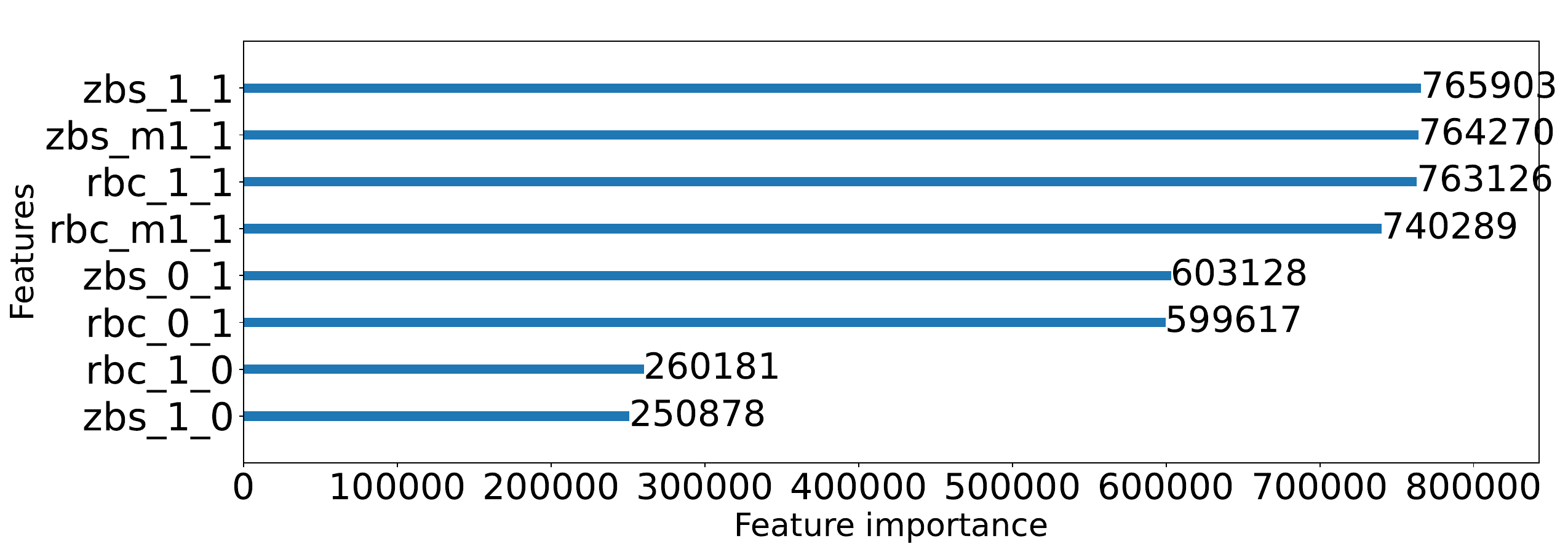}
\includegraphics[trim=1.0cm 1.1cm 0.8cm 1.1cm, clip, width=0.31\textwidth]{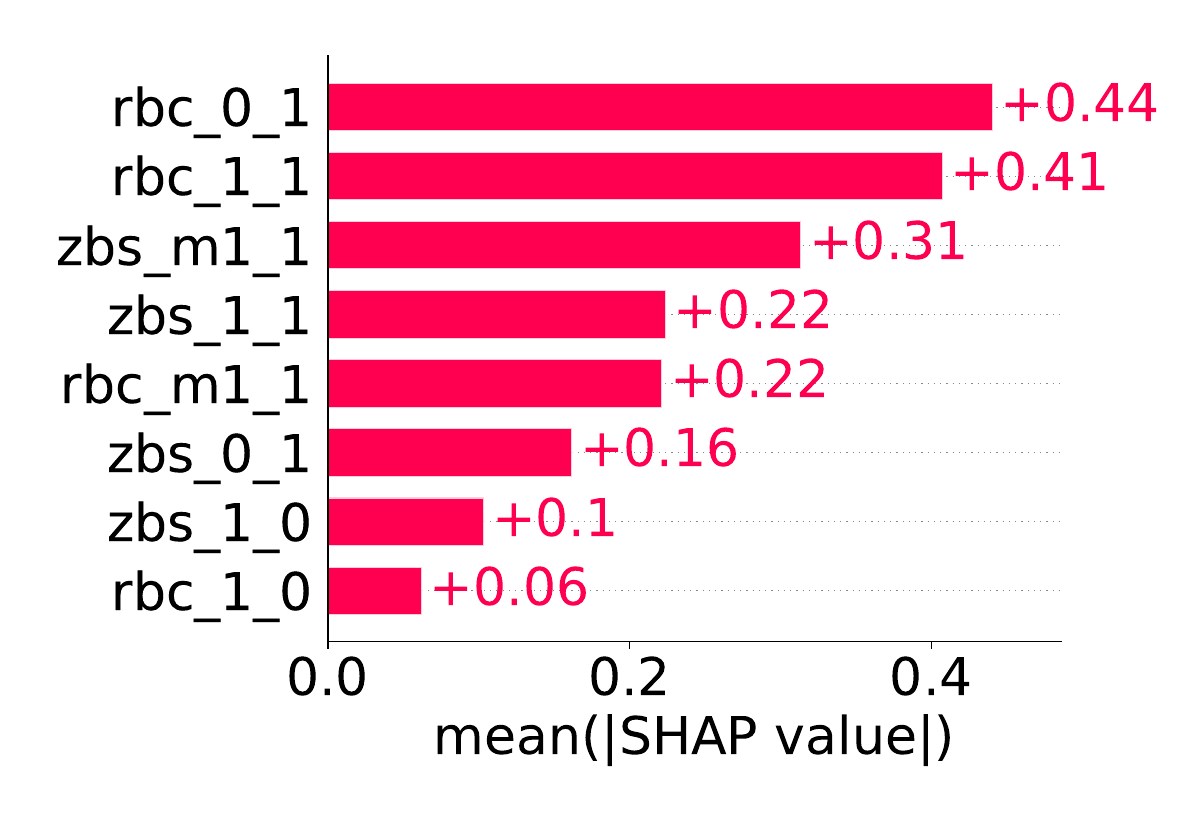}
\caption{Comparison of feature importance to quasisymmetry between LightGBM (top) and LightGBM LSS (bottom) models.}
\label{importance_regressors}
\end{figure}

\begin{figure}[ht]
\centering
\includegraphics[trim=0.3cm 0.3cm 0.7cm 1.1cm, clip, width=0.5\textwidth]{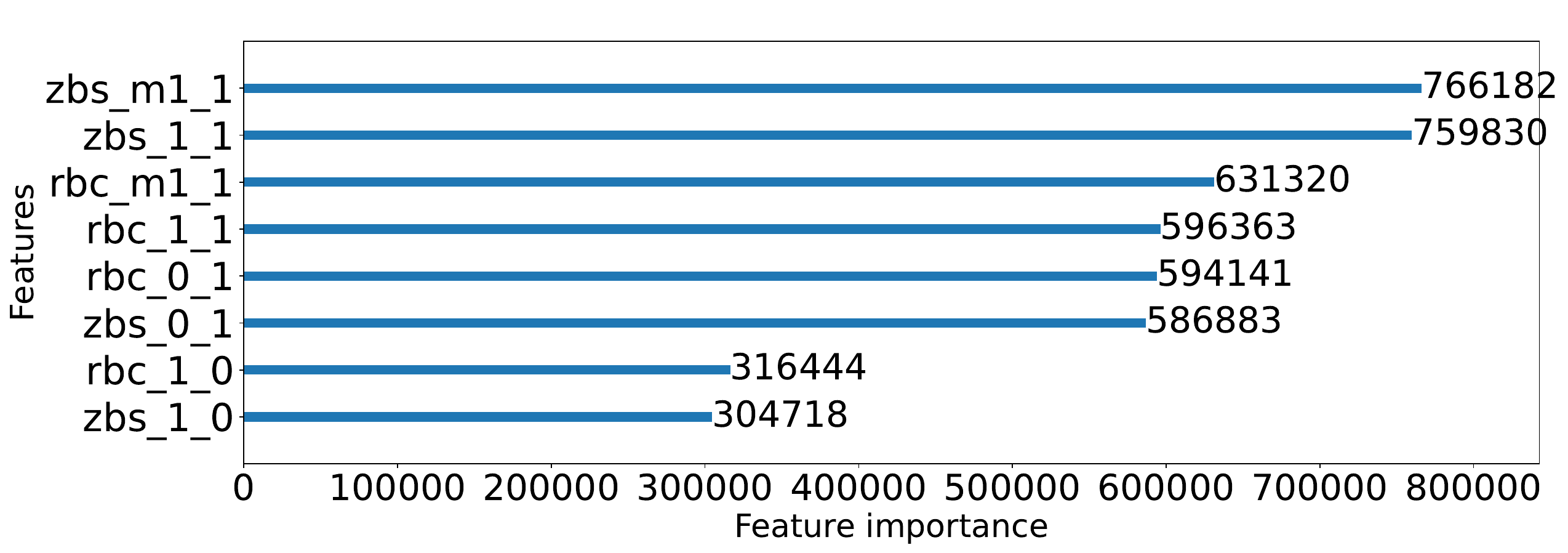}
\includegraphics[trim=1.0cm 1.1cm 0.8cm 1.1cm, clip, width=0.31\textwidth]{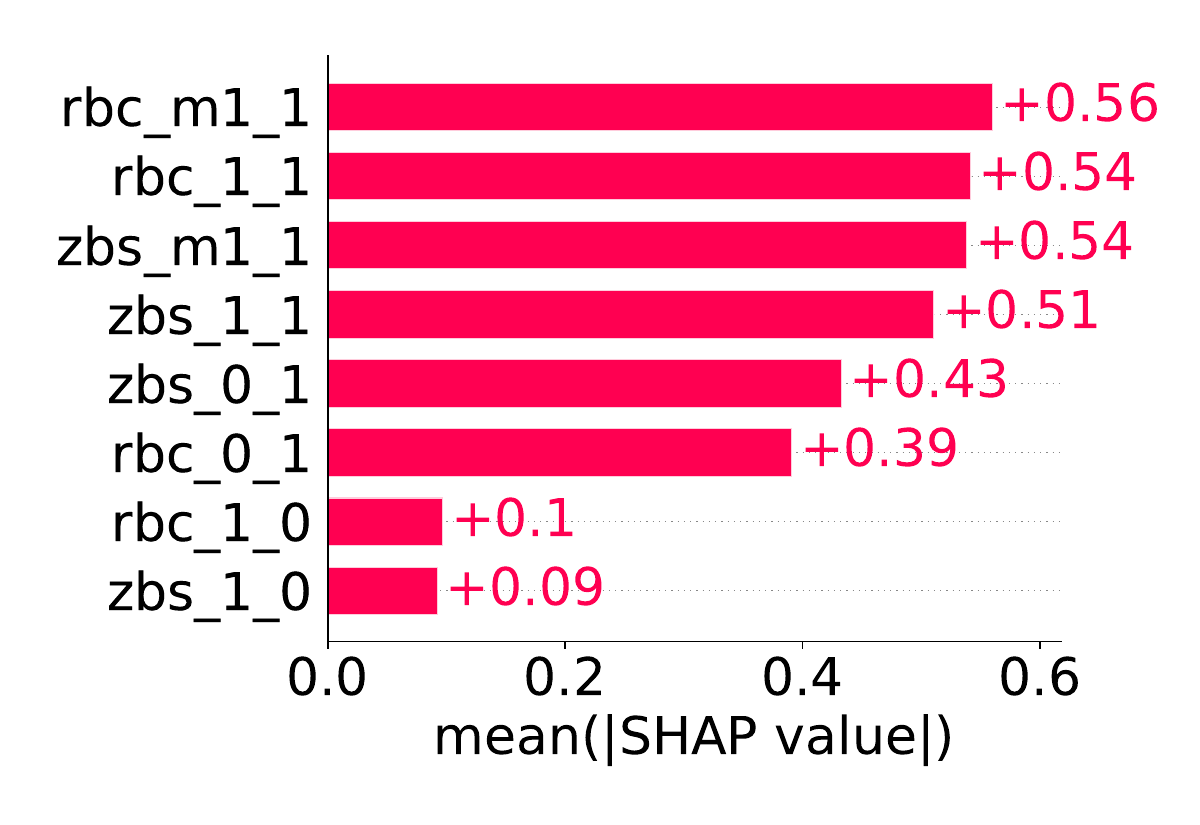}
\caption{Comparison of feature importance to quasi-isodynamic between LightGBM (top) and LightGBM LSS (bottom) models.}
\label{qi_importance_regressors}
\end{figure}

In LightGBM, the feature importance represents how often each feature was used to split the data in decision trees during training.
In LightGBM LSS, SHAP values help decompose each instance's prediction by quantifying each feature's contribution to the prediction.
This is achieved by comparing predictions when each feature is included versus excluded, considering interactions with other features to attribute contributions. For the non-probabilistic version (\cref{importance_regressors}, left), $\text{ZBS}_{1,1}$, $\text{ZBS}_{-1,1}$, $\text{RBC}_{1,1}$, and $\text{RBC}_{-1,1}$ were identified as the most influential parameters regarding quasisymmetry.
However, the LightGBM LSS model indicated that $\text{RBC}_{0,1}$, $\text{RBC}_{1,1}$, $\text{ZBS}_{-1,1}$, and $\text{RBC}_{-1,1}$ were the most significant (\cref{importance_regressors}, right).
The LightGBM model (\cref{qi_importance_regressors}, left), applied to the quasi-isodynamic data revealed that $\text{ZBS}_{-1,1}$ and $\text{ZBS}_{1,1}$ were the most influential Fourier coefficients on this output and LightGBM LSS highlights the importance of $\text{RBC}_{-1,1}$, $\text{RBC}_{1,1}$, $\text{ZBS}_{-1,1}$ and $\text{ZBS}_{1,1}$ (\cref{qi_importance_regressors}, right).
However, using these models, we are only able to identify the most significant features for each Gaussian in the mixture individually.
Consequently, \cref{importance_regressors} (bottom) and \cref{qi_importance_regressors} (bottom) show the influence of the Fourier coefficients for the first Gaussian in each mixture, specifically the one with the lowest quasisymmetry and quasi-isodynamic.

\begin{figure}[ht]
    \centering
    \includegraphics[trim=0cm 0.0cm 0cm 0.0cm, clip, width=0.45\textwidth]{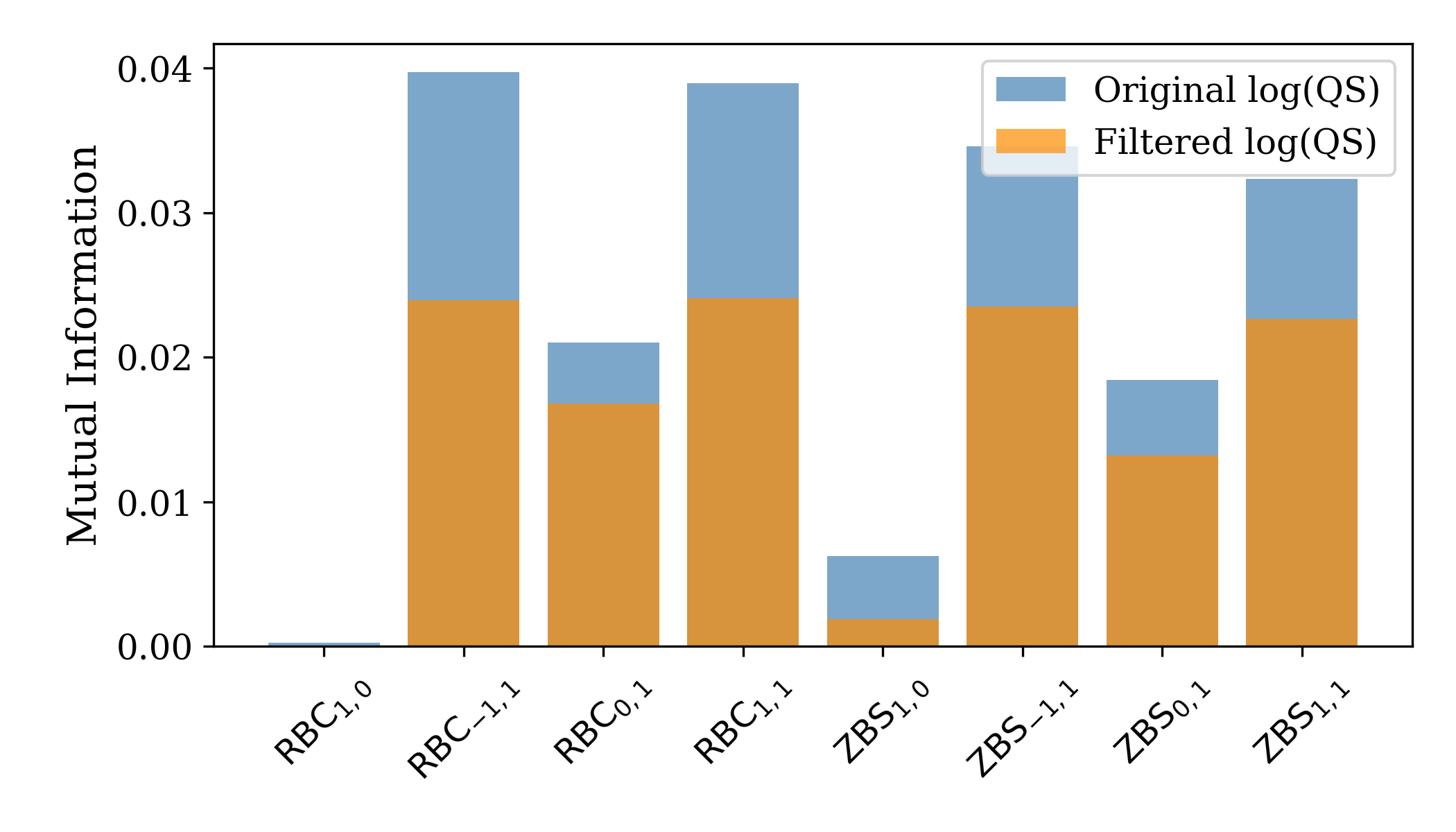}
    \includegraphics[trim=0cm 0.0cm 0cm 0.0cm, clip, width=0.45\textwidth]{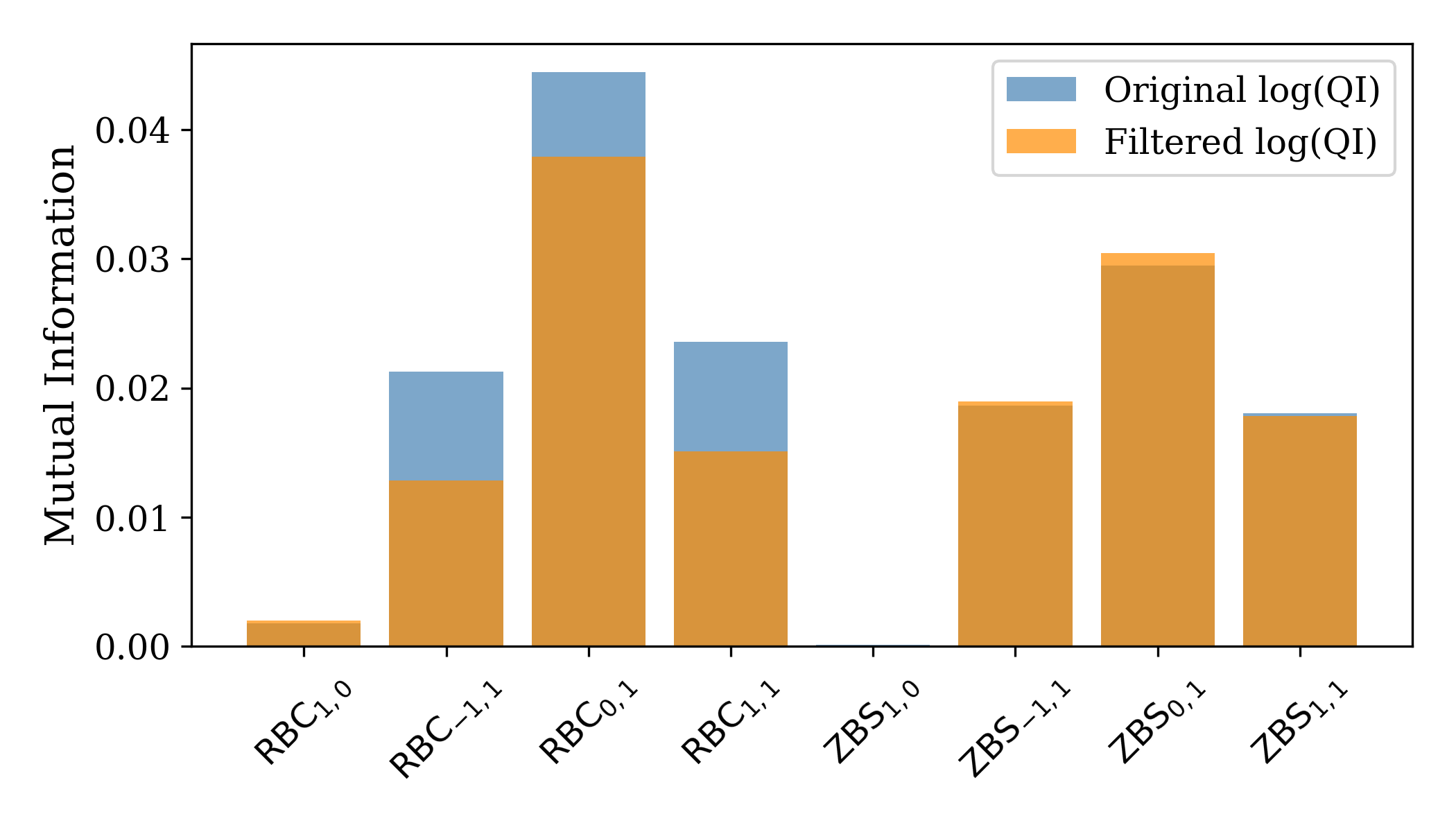}
    \caption{Feature importance of the Fourier coefficients of the plasma boundary on quasisymmetry (top) and quasi-isodynamicity (bottom).}
    \label{fig:mutual_importance}
\end{figure}

We corroborate the results found here using a Mutual Information algorithm \cite{Estevez2009}, shown in \cref{fig:mutual_importance}.
This is a non-parametric measure of the dependence between variables.
Each bar in \cref{fig:mutual_importance} corresponds to a specific spectral coefficient from the boundary representation, labeled in the format $\text{RBC}_{m,n}$ or $\text{ZBS}_{m,n}$, and its height indicates how much information it provides about quasisymmetry (top) and quasi-isodynamicity (bottom).
Both filtered and unfiltered data of outliers are shown.
As predicted by the gradient boosting models, the parameters $\text{RBC}_{1,0}$ or $\text{ZBS}_{1,0}$ have a small impact on omnigenity.
The importance of the remaining parameters is in line with the Gradient Boosting approaches.
To gain a more comprehensive understanding of feature importance across the entire distribution, in the next section, SHAP analysis is used in the context of feed-forward neural networks.

\section{Feed-Forward Neural Network}
\label{section:FFNN}

In this section, we train a neural network in order to identify the most influential Fourier coefficients affecting the overall quasisymmetry and quasi-isodynamic distributions. 
Additionally, we assess whether a relatively simple neural network model can effectively fit the data, resulting in lower prediction errors compared to the LightGBM models. 

The fundamental unit of a neural network is the neuron, which acts as a computational element.
Neurons are interconnected through weights, which control the strength of these connections. Each neuron receives input signals, calculates a weighted sum of these inputs, and applies an activation function to produce its output.
This output is then forwarded to neurons in subsequent layers.
In this work, a feed-forward neural network (FFNN) is used \cite{Svozil1997}.
FFNNs are a class of artificial neural networks characterized by a unidirectional flow of information, without any feedback loops.
%
%
%


For modeling quasisymmetry properties, a multi-layer feed-forward neural network architecture is employed.
%
This network starts with an input layer of 8 neurons, corresponding to the Fourier coefficients. However, it features five hidden layers: the first layer includes 512 neurons, the second 256 neurons, the third 128 neurons, the fourth 64 neurons, and the fifth 16 neurons.
In contrast, for modeling quasi-isodynamic properties, a different feed-forward neural network was designed.
The network architecture consists of five hidden layers: the first with 512 neurons, the second with 256 neurons, the third with 128 neurons, the fourth with 64 neurons, and the fifth with 16 neurons.
The output layer comprises a single neuron with a linear activation function.

\begin{figure}
\centering
\includegraphics[trim=0.6cm 0.1cm 3.3cm 2.5cm, clip, width=0.47\textwidth]{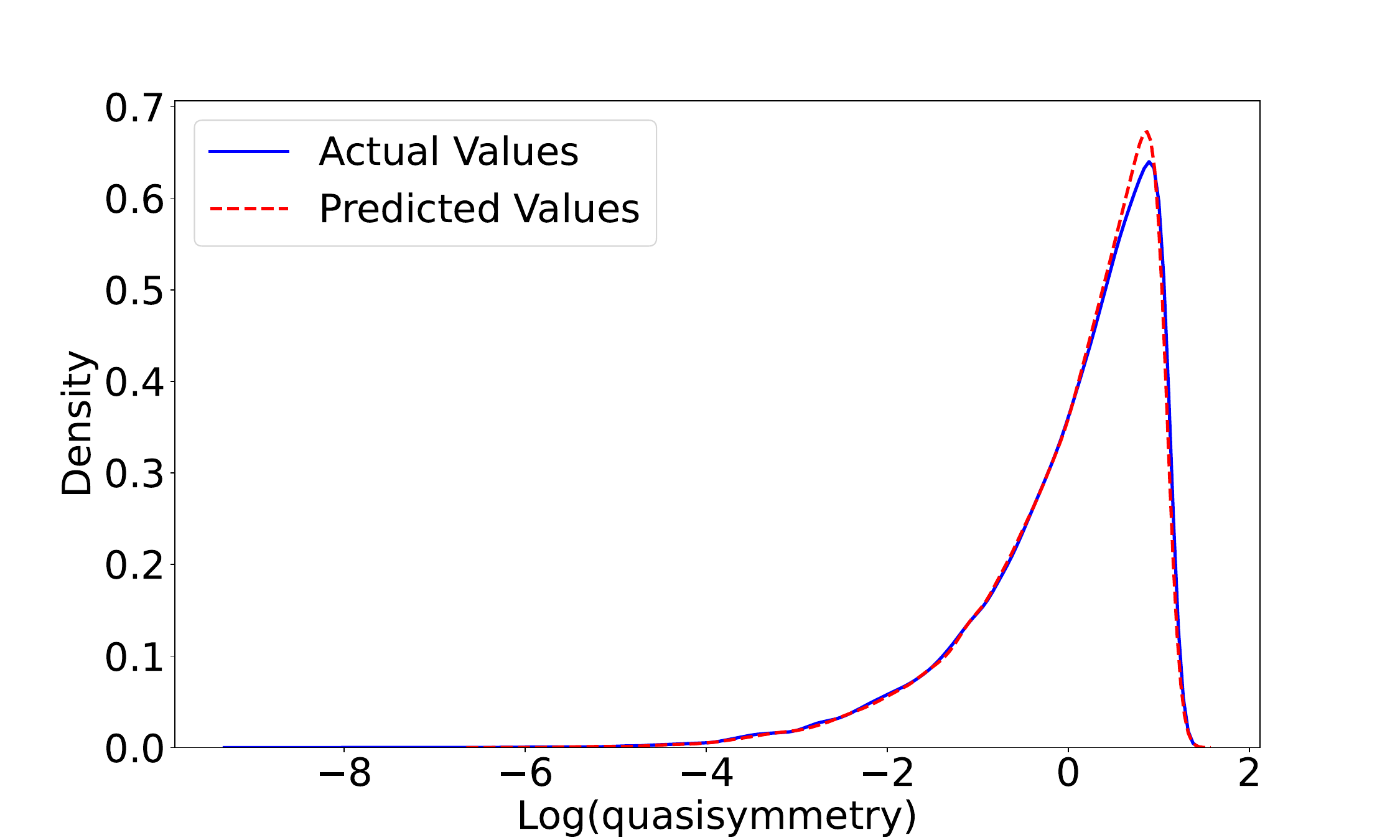}
\includegraphics[trim=0.9cm 0.9cm 0.6cm 0.7cm, clip, width=0.42\textwidth]{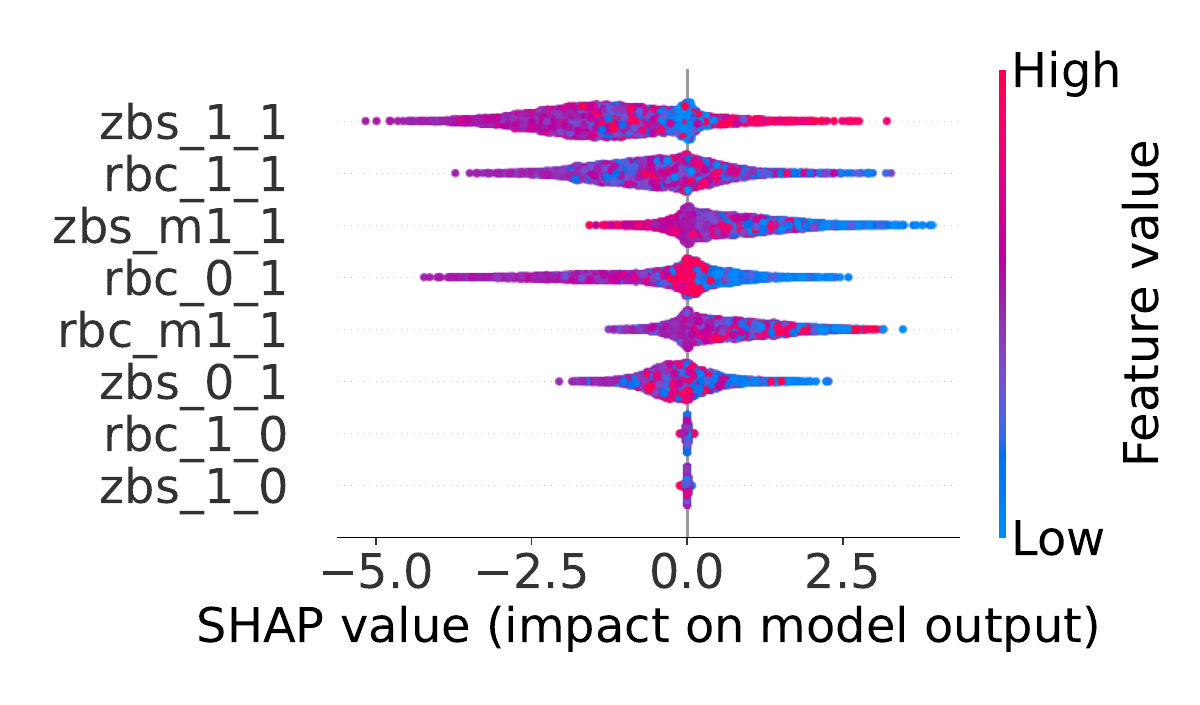}
\caption{Comparison of actual and predicted distributions of quasisymmetry using a feedforward neural network (top) and most important features according to their impact on quasisymmetry (bottom).}
\label{fig:nn_qs}
\end{figure}

\begin{figure}
\centering
\includegraphics[trim=0.6cm 0.1cm 3.3cm 2.5cm, clip, width=0.47\textwidth]{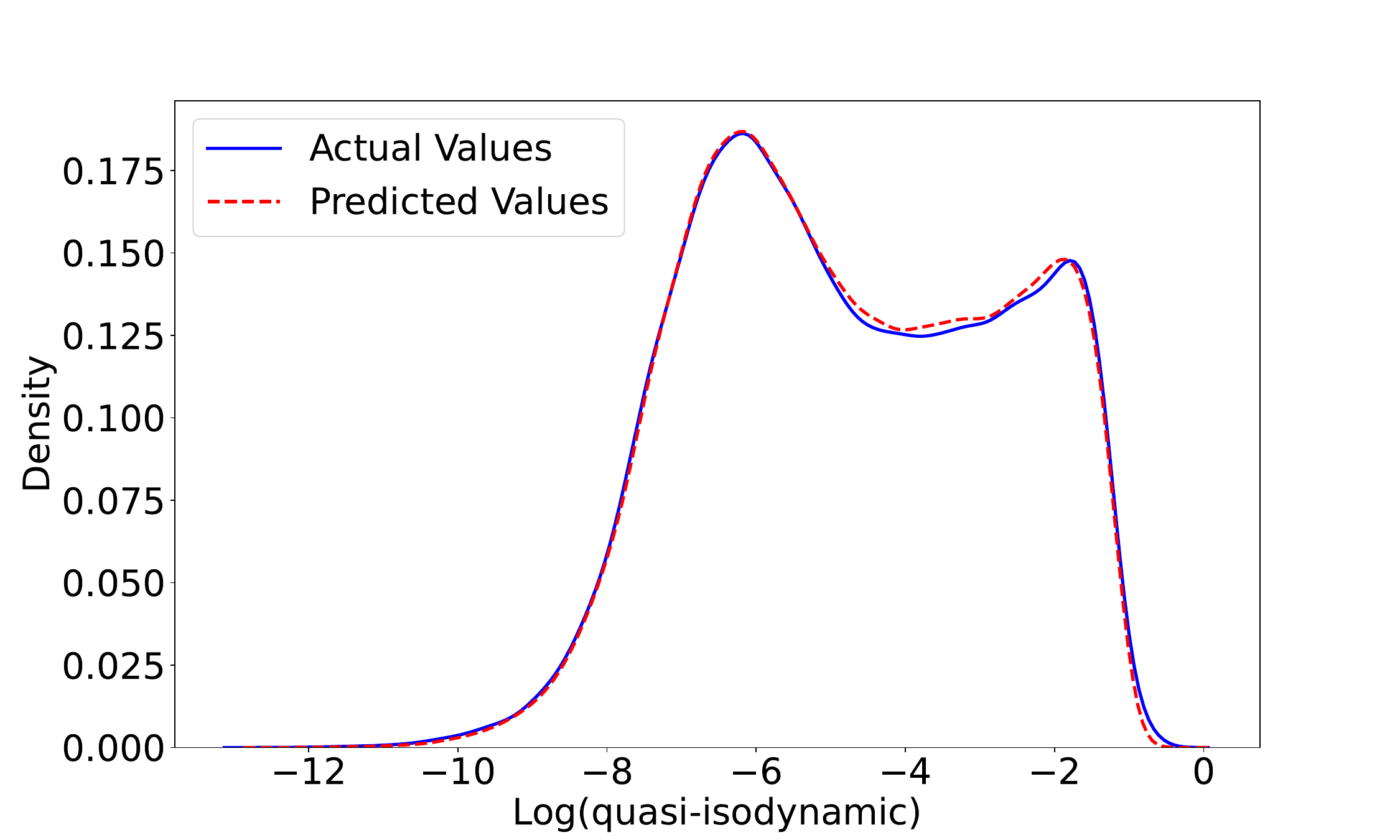}
\includegraphics[trim=0.9cm 0.9cm 0.6cm 0.7cm, clip, width=0.42\textwidth]{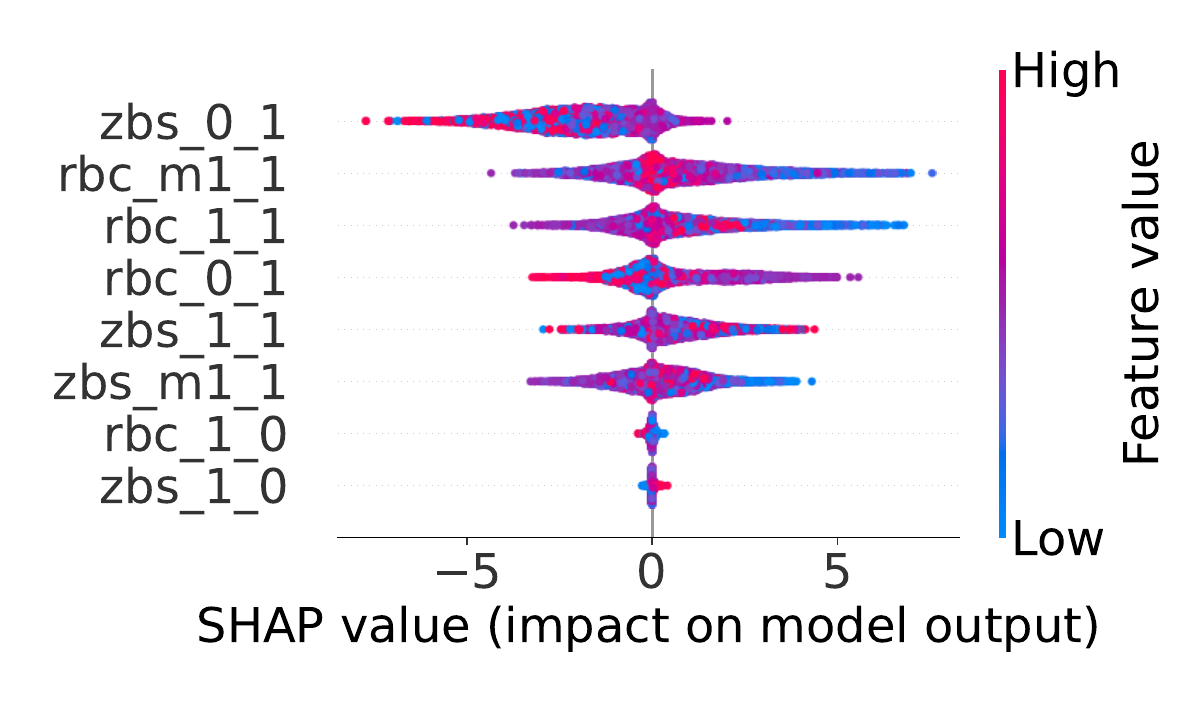}
\caption{Comparison of actual and predicted distributions of quasi-isodynamic using a feedforward neural network (top) and most important features according to their impact on quasi-isodynamic (bottom).}
\label{fig:nn_qi}
\end{figure}

The Adam optimizer was used for training both models, with a learning rate set at 0.001.
The networks were trained over 500 epochs with a batch size of 500.
A validation split of 20\% was applied. The performance of the two neural network models is assessed using statistical measures, namely the $R^{2}$ score and the MSE.
For the quasi-isodynamic neural network, the $R^{2}$ score of 0.9909 suggests that the model explains approximately 99.09\% of the data's variance, demonstrating a high degree of predictive accuracy.
In \cref{fig:nn_qi} (top), we compare the predicted and actual distributions for this model, showing very good agreement. 
The MSE for this model is 0.0384.
In comparison, the quasisymmetry neural network also achieved an $R^{2}$ score of 0.9909. 
In \cref{fig:nn_qs} (top), we compare predicted and actual distributions for the quasisymmetry model.
While this model yields a lower MSE of 0.0092 when compared with the QI one, the predicted distribution agrees well with the actual distribution.

Finally, we use SHAP plots created using the SHAP Python library \cite{Lundberg2017APredictions} to analyze the impact of individual features on the model's output for quasi-isodynamic and quasisymmetry properties.
We employ a model-agnostic Kernel SHAP method, where the Shapley value for a feature represents the average marginal contribution of that feature to the prediction, computed across all possible feature subsets.
Essentially, the distribution of SHAP values shows how changes in these features lead to variations in the model's output.
In our case, the SHAP explainer was instantiated with a single-sample background drawn from the training data to approximate the conditional expectations required for the Shapley computation.
For quasi-isodynamicity, the SHAP summary plot in \cref{fig:nn_qi} (bottom) highlights $\text{ZBS}_{0,1}$, $\text{RBC}_{-1,1}$, and $\text{RBC}_{1,1}$ as the most important features affecting this distribution. 
Similarly, for quasisymmetry, the SHAP summary plot in \cref{fig:nn_qs} (bottom) identifies the features that most significantly influence the model's predictions. The plot uses color coding to represent feature values, showing that features such as $\text{ZBS}_{1,1}$, $\text{RBC}_{1,1}$, and $\text{ZBS}_{-1,1}$ have a strong impact on the model's output. 
As obtained with the previous models, $\text{RBC}_{1,0}$ and $\text{ZBS}_{1,0}$ are identified as Fourier coefficients with little to no impact on omnigenity.

\section{Conclusions}

In this work, we use several techniques to create a database of stellarator configurations, obtain insights from it, and use it to train new models.
The focus in this work was to identify which stellarator configurations exhibit the most desirable values for quasisymmetry and quasi-isodynamicity.
Nevertheless, geometric properties of the equilibrium magnetic fields obtained, such as elongation, aspect ratio, magnetic well, and rotational transform, are added to the database and analyzed.
We find that the highest levels of correlation are found between quasisymmetry and mirror ratio.
Pareto fronts for different parameter pairs are shown, and dimensionality reduction techniques are applied to obtain models that depend on only two parameters instead of several Fourier coefficients of the plasma boundary. Specifically, we apply a supervised autoencoder approach which appears to be a promising avenue as an efficient optimization strategy.
Then, gradient boost models such as LightGBM and LightGBM LSS are trained on the original dataset.
We find that while LightGBM LSS yields a marginal distribution function more similar to the one in the data, LightGBM predictions exhibited fewer and smaller errors throughout the entire logarithm of the quasisymmetry and quasi-isodynamic distributions.
For quasisymmetry, the errors were larger at lower quasisymmetry values, likely due to the database's scarcity of low quasisymmetry values, and the opposite happened for quasi-isodynamic, possibly due to its double-peaked character.
Additionally, a feed-forward neural network model is trained, which outperformed the tree-based models in predicting each studied parameter, suggesting that the feature importance derived from the neural networks may be more accurate.
Across all models, the two Fourier coefficients corresponding to the $\text{RBC}_{1,0}$ and $\text{ZBS}_{1,0}$ components were consistently identified as the least important features.
Finally, a classification model was employed to evaluate whether VMEC can converge and whether quasisymmetry has a value below 10.
This model achieves a total accuracy of 99.2\%.

The models obtained here, such as the gradient boosting models and the feed-forward neural networks, can be used as fast surrogate models for the evaluation of the prediction properties of stellarators to be included in the optimization of new devices.
Furthermore, the model reduction techniques employed here, such as autoencoders, can be used to make such optimization more efficient by optimizing in the reduced two-dimensional space instead of the Fourier coefficients of the boundary.
Future work includes the use of variational autoencoders and deeper networks to obtain a dimensionality reduction with better reproduction capabilities.
Additionally, while our main focus was on omnigenity metrics, this study can be performed on other performance metrics that are present in the database, such as MHD stability via the magnetic well, or from additional metrics such as non-linear heat fluxes to evaluate turbulent transport.
A more encompassing database with various field periods and more compact stellarators will be created in the future.

\section*{Acknowledgements}

R. J. is supported by the Portuguese FCT-Fundação para a Ciência e Tecnologia, under Grant 2021.02213.CEECIND and DOI  \href{https://doi.org/10.54499/2021.02213.CEECIND/CP1651/CT0004}{10.54499/2021.02213.CEECIND/CP1651/CT0004}.
R. J was also supported by the National Science Foundation under Grant No. 2409066.
This work has been carried out within the framework of the EUROfusion Consortium, funded by the European Union via the Euratom Research and Training Programme (Grant Agreement No 101052200– EUROfusion). Views and opinions expressed are however those of the author(s) only and do not necessarily reflect those of the European Union or the European Commission. Neither the European Union nor the European Commission can be held responsible for them.
IPFN activities were supported by FCT - Fundação para a Ciência e Tecnologia, I.P. by project reference UIDB/50010/2020 and DOI  \href{https://doi.org/10.54499/UIDB/50010/2020}{10.54499/UIDB/50010/2020}, by project reference UIDP/50010/2020 and DOI \href{https://doi.org/10.54499/UIDP/50010/2020}{10.54499/UIDP/50010/2020} and by project reference LA/P/0061/202 and  DOI \href{https://doi.org/10.54499/LA/P/0061/2020}{10.54499/LA/P/0061/2020}.
This work used the Jetstream2 at Indiana University through allocation PHY240054 from the Advanced Cyberinfrastructure Coordination Ecosystem: Services \& Support (ACCESS) program which is supported by National Science Foundation grants \#213859, \#2138286, \#2138307, \#2137603 and \#2138296.
This research used resources of the National Energy Research
Scientific Computing Center, a DOE Office of Science User Facility
supported by the Office of Science of the U.S. Department of Energy
under Contract No. DE-AC02-05CH11231 using NERSC award
NERSC DDR-ERCAP0030134.
This research used resources of the Oak Ridge Leadership Computing Facility at the Oak Ridge National Laboratory, which is supported by the Office of Science of the U.S. Department of Energy under Contract No. DE-AC05-00OR22725.


\bibliographystyle{unsrt}
\bibliography{references_mendeley}

\end{document}